\documentclass[longauth]{aa} 
\usepackage{xcolor}
\usepackage{support-caption}
\usepackage{subcaption}
\usepackage{graphicx}
\usepackage{threeparttable}
\usepackage{float}
\usepackage{booktabs}
\usepackage{longtable}
\usepackage{lscape}
\usepackage{amsmath} % Required for \substack
\usepackage{arydshln}

\newcommand{\ha}{\ifmmode {\rm H}\alpha \else H$\alpha$\fi}
\newcommand{\hb}{\ifmmode {\rm H}\beta \else H$\beta$\fi}
\newcommand{\lya}{\ifmmode {\rm Ly}\alpha \else Ly$\alpha$\fi}
\newcommand{\pg}{\ifmmode {\rm P}\gamma \else Pa$\gamma$\fi}
\newcommand{\lyb}{\ifmmode {\rm Ly}\beta \else Ly$\beta$\fi}
\newcommand{\lyg}{\ifmmode {\rm Ly}\gamma \else Ly$\gamma$\fi}

\newcommand{\flyc}{\ifmmode \mathrm{f}_\mathrm{esc}\mathrm{(LyC)} \else $\mathrm{f}_\mathrm{esc}\mathrm{(LyC)}$\fi}
\newcommand{\lime}{\textsc{LiMe}\,}

\def\ergs{\ifmmode \mathrm{erg\hspace{1mm}s}^{-1} \else erg s$^{-1}$\fi}

\def\micron{\ifmmode \mu\mathrm{m} \else $\mu$m\fi}
\def\msun{\ifmmode \mathrm{M}_{\odot} \else M$_{\odot}$\fi}
\def\msunyr{\ifmmode \mathrm{M}_{\odot} \hspace{1mm}{\rm yr}^{-1} \else $\mathrm{M}_{\odot}$ yr$^{-1}$\fi}
\def\zsun{\ifmmode Z_{\odot} \else Z$_{\odot}$\fi}
\def\lsun{\ifmmode L_{\odot} \else L$_{\odot}$\fi}
\def\mstar{\ifmmode \mathrm{M}_{\star} \else M$_{\star}$\fi}

\newcommand{\jwst}{JWST}

\newcommand{\NIRSpec}{NIRSpec}
\newcommand{\NIRCam}{NIRCam}

\usepackage{txfonts}
\newcommand{\orcid}[1]{\href{https://orcid.org/#1}{\includegraphics[width=10pt]{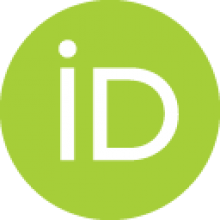}}}
\usepackage{hyperref}
\hypersetup{colorlinks, % = true per default
 linkcolor = blue,
 urlcolor = blue,
 citecolor = blue,
 anchorcolor = blue
}
\begin{document}

\title{The AGN nature of strong \ion{C}{III}]  emitters in the Early Universe with JWST}

%Unraveling, Exploring, Unveiling, Deciphering, Probing, Untangling

 \subtitle{}
     \author{F. Arevalo-Gonzalez \orcid{0009-0005-5968-8553}
 \inst{1,2,3}
\and R. Tripodi \orcid{0000-0002-9909-3491}
 \inst{1}
 \and M. Llerena \orcid{0000-0003-1354-4296}
\inst{1}
 \and L. Pentericci \orcid{0000-0001-8940-6768}
 \inst{1}
 \and A. Plat \orcid{0000-0003-0390-0656}
 \inst{4}
 \and G. Barro \orcid{0000-0001-6813-875X}
 \inst{5}
 % People who gave comments
 \and R. O. Amorín \orcid{0000-0001-5758-1000}
 \inst{6}
 \and B. Backhaus \orcid{0000-0001-8534-7502}
 \inst{7}
 \and A. Calabrò \orcid{0000-0003-2536-1614}
 \inst{1}
 \and N. J. Cleri \orcid{0000-0001-7151-009X}
 \inst{8,9,10}
 \and M. Dickinson \orcid{0000-0001-5414-5131}
 \inst{11}
 \and J. S. Dunlop \orcid{0000-0002-1404-5950}
 \inst{12}
 \and S. L. Finkelstein \orcid{0000-0001-8519-1130}
 \inst{13,14}
 \and M. Giavalisco \orcid{0000-0002-7831-8751}
 \inst{15}
 \and M. Hirschmann \orcid{0000-0002-3301-3321}
 \inst{4}
 \and J. Kartaltepe \orcid{0000-0001-9187-3605}
 \inst{16}
 \and A. M. Koekemoer \orcid{0000-0002-6610-2048}
 \inst{17}
 \and R. A. Lucas \orcid{0000-0003-1581-7825}
 \inst{17}
 \and L. Napolitano \orcid{0000-0002-8951-4408}
 \inst{1}
 \and E. Piconcelli \orcid{0000-0001-9095-2782}
 \inst{1}
 \and A. J. Taylor \orcid{0000-0003-1282-7454}
 \inst{13,14}
 \and F. Tombesi \orcid{0000-0002-6562-8654}
 \inst{3,1,18}
 \and J. R. Trump \orcid{0000-0002-1410-0470}
 \inst{19}
 % further people who participate to the telecon and will give more comments
 %\inst{1}
 \and X. Wang \orcid{0000-0002-9373-3865}
 \inst{20,21,22}
 }
 \institute{\textit{INAF – Osservatorio Astronomico di Roma, via Frascati 33, 00078, Monteporzio Catone, Italy}\\ %1
 \email{flor.arevalogonzalez@inaf.it}
 \and 
 \textit{Dipartimento di Fisica, Università di Roma Sapienza, Città Universitaria di Roma - Sapienza, Piazzale Aldo Moro, 2, 00185, Roma, Italy} %2
 \and
 \textit{Physics Department, Tor Vergata University of Rome, Via della Ricerca Scientifica 1, 00133 Rome, Italy} %3
  \and
 \textit{Institute of Physics, Laboratory for Galaxy Evolution and Spectral Modelling, EPFL, Observatoire de Sauverny, Chemin Pegasi 51, 1290 Versoix, Switzerland} %4
 \and
 \textit{University of the Pacific, Stockton, CA 90340 USA} %5
 \and
 \textit{Instituto de Astrofísica de Andalucía (IAA-CSIC), Glorieta de la Astronomía s/n, 18008 Granada, Spain} %6
 \and
 \textit{Department of Physics and Astronomy, University of Kansas, Lawrence, KS 66045, USA} %7
\and 
 \textit{Department of Astronomy and Astrophysics, The Pennsylvania State University, University Park, PA 16802, USA} %8
 \and
\textit{Institute for Computational and Data Sciences, The Pennsylvania State University, University Park, PA 16802, USA} %9
\and
\textit{Institute for Gravitation and the Cosmos, The Pennsylvania State University, University Park, PA 16802, USA} %10
\and
\textit{NSF’s NOIRLab, Tucson, AZ 85719, USA}%11
\and
\textit{Institute for Astronomy, University of Edinburgh, Royal Observatory, Edinburgh, EH9 3HJ, UK} %12
\and
\textit{Department of Astronomy, The University of Texas at Austin, Austin, TX, USA} %13
 \and
 \textit{Cosmic Frontier Center, The University of Texas at Austin, Austin, TX, USA} %14
 \and
 \textit{University of Massachusetts Amherst, 710 North Pleasant Street, Amherst, MA 01003-9305, USA} %15
 \and
 \textit{Laboratory for Multiwavelength Astrophysics, School of Physics and Astronomy, Rochester Institute of Technology, 84 Lomb Memorial Drive, Rochester, NY
14623, USA} %16
 \and
 \textit{Space Telescope Science Institute, 3700 San Martin Drive, Baltimore, MD 21218, USA} %17
\and
\textit{INFN - Rome Tor Vergata, Via della Ricerca Scientifica 1, 00133 Rome, Italy} %18
 \and
 \textit{Department of Physics, 196A Auditorium Road, Unit 3046, University of Connecticut, Storrs, CT 06269, USA} %19
 \and
 \textit{School of Astronomy and Space Science, University of Chinese Academy of Sciences (UCAS), Beijing 100049, China} %20
 \and
 \textit{National Astronomical Observatories, Chinese Academy of Sciences, Beijing 100101, China} %21
 \and
 \textit{Institute for Frontiers in Astronomy and Astrophysics, Beijing Normal University, Beijing 102206, China} %22
}

%\date{Accepted XXX. Received YYY; in original form ZZZ}

%273 words -- maximum 300
\abstract{The semi-forbidden \ion{C}{III}] $\lambda\lambda$1907,1909 doublet is a key tracer of high-ionization emission in the early universe. We present a study of \ion{C}{III}] emission in galaxies at z = 5–7, using publicly available JWST/NIRSpec prism data from programs including CEERS, JADES, and RUBIES, along with data from the ongoing CAPERS survey. We built a sample of 61 \ion{C}{III}]-emitting galaxies, and we classified them as star-forming or active galactic nuclei (AGN) host galaxies using (1) rest-frame UV and optical emission-line diagnostic diagrams, and (2) the presence/absence of broad Balmer emission lines. The UV diagnostics are based on the combination of the rest-frame equivalent width (EW) of \ion{C}{III}] versus the ratio  \mbox{\ion{C}{III}]/\ion{He}{II} $\lambda$1640}, and the EW of \ion{C}{IV} versus the ratio \mbox{\ion{C}{IV}/\ion{He}{II} $\lambda$1640}. For optical diagnostics, we employ the OHNO diagram, which combines four optical emission lines that span a range in ionization potential — [\ion{O}{III}] $\lambda$5007, H$\beta$, [\ion{Ne}{III}] $\lambda$3869, and [\ion{O}{II}] $\lambda\lambda$3727,3729- and we find it has a low efficiency on separating AGN from SFG. We find that half of the sources in our sample (29 out of 61 galaxies) exhibit at least one secure indication of AGN activity while 13 are potential AGNs based on the \ion{C}{III}] diagnostic. Physical properties, including stellar mass and star formation rate, are derived through spectral energy distribution modeling with \textsc{Bagpipes}. Our analysis reveals that JWST is uncovering a population of strong \ion{C}{III}] emitters at high redshift ($5 < z < 7$) with a median EW of {22.8~\AA} (16th--84th percentile: {12.5--51.5~\AA}). This EW is higher than that of a control sample of \ion{C}{III}] emitters at intermediate redshift ($3 < z < 4$) with a median EW of 4.7~\AA. We find that for the same range of M$_{\mathrm{UV}}$, the \ion{C}{III}] EW increases by $\sim${0.67~dex} from $3 < z < 4$ to $5 < z < 7$, indicating strong redshift evolution in the line's strength. We also find that \ion{C}{III}] emitters span the full range of the star-forming main sequence, where the majority of galaxies below the sequence have high \ion{C}{III}] EW. Finally, we identify five sources in our sample as Little Red Dots (LRDs) according to the selection criteria in the literature; while four of these have already been identified as LRD in the literature, one is presented here for the first time.}

% 5 {} token are mandatory
 
 %\abstract


 \keywords{galaxies: high-redshift, cosmology: dark ages, reionization, first stars}

 \maketitle

 %List of useful commands:
 %\lya
 %\citep[see also][]{xu2022, izotov2022} o \citep{Planck2020}
 %Sec.~\ref{sec:data}
 %\texttt{Spec2Pipeline} for codes
 %\section{Introduction} \label{sec:intro}
 %\subsection{Measurements of physical parameters }\label{sec:SED_fitting}
 
 % Iserire figura:
 %\begin{figure}[ht!]
 %\includegraphics[width=\linewidth]{images/O3hb_mass.pdf}
 %\caption{
 %\label{fig:AGN}}
 %\end{figure}
%-------------------------------------------------------------------

%\newpage

\section{Introduction} \label{sec:intro}

The Epoch of Reionization (EoR) represents a pivotal phase in cosmic history, following the initial "cosmic dawn". During this era, ionizing radiation from the first stars and galaxies, and potentially from the earliest active galactic nuclei (AGN), ended the cosmic dark ages and drove the transition of the intergalactic medium (IGM) from neutral to ionized at $z\gtrsim5.3$ \citep[e.g.,][]{Bosman_reionization, 2024A&A_Spina, 2015_Becker}. Studying the sources responsible for this process—whether primordial starbursts or AGN—is crucial for understanding the drivers of reionization \citep[e.g.,][]{2024JCAP_Hegde}. However, observing these distant sources presents significant challenges. Emission lines provide a powerful window into the physical conditions of galaxies, encoding information about their ionization state, gas density, metallicity, and radiation field. During the EoR, observational access is largely limited to rest-frame UV diagnostics, making the choice of emission lines particularly important. While Ly$\alpha$ is often the brightest UV line, its resonant nature causes it to be heavily attenuated and scattered by neutral hydrogen in the IGM, complicating its interpretation as a probe of intrinsic galaxy properties. We therefore focus on alternative UV lines such as \ion{C}{III}] $\lambda\lambda$1907,1909, which arises from nebular gas and provides more direct constraints on the physical conditions within high-redshift galaxies \citep[e.g.,][]{LeFevre2019}.

The semi-forbidden doublet of doubly-ionized carbon, \ion{C}{III}] $\lambda\lambda$1907,1909 (hereafter \ion{C}{III}]), offers a powerful opportunity to reveal AGN hosts and star-forming galaxies (SFGs) at high-z. Indeed, it is typically the second brightest UV emission line after Ly$\alpha$, making it readily observable. Additionally, \ion{C}{III}] traces gas with high-energy radiation, as it requires photons with energies between $\sim$24.4 eV and 47.9 eV (the range to create and excite C$^{2+}$). This requires intense radiation fields typically associated with either AGN or young, metal-poor starburst galaxies.

Extensive pre-James Webb Space Telescope \citep[\jwst,][]{Gardner2023} observations of galaxies at $z \sim 2$–4, using both ground-based spectroscopy and Hubble Space Telescope at low redshift, have established the diagnostic power of rest-frame UV emission lines for identifying ionization sources and constraining galaxy physical conditions. Studies have shown that \ion{C}{III}] is sensitive to metallicity, ionization parameter, and the hardness of the ionizing radiation field in star-forming galaxies, with particularly strong emission observed in young, metal-poor systems \citep[e.g.,][]{Stark2014, p12_nakajima2018}. Recently  \citet{Cunningham_2024} analysed  a sample of 62 sources showing  \ion{C}{III}] emission (as well as  Ly$\alpha$) at $3 < z < 4$ from the VANDELS survey: 
these sources have rest-frame EW(\ion{C}{III}]) between $\simeq$ 1-15 \AA, with   a median of  4.7 \AA. The study showed that only a small fraction of the sources (10 out of 62, i.e. around 14\%) were identified as AGN, with rest being dominated by star formation \citep[see also][]{LeFevre2019}.
\\
These pre-JWST results demonstrate that UV emission-line diagnostics provide a robust framework for distinguishing between star formation and AGN ionization, motivating their application to JWST observations of galaxies in the reionization era.

\jwst~has fundamentally transformed our ability to study these \ion{C}{III}] emitters at high-z. For the first time, we have access to a powerful, sensitive infrared instrument capable of efficiently observing simultaneously the rest-frame ultraviolet and optical spectra of high-redshift galaxies. As the \ion{C}{III}] feature is redshifted into the near-infrared for objects during the EoR, JWST's instruments are ideally suited for its detection and characterization. JWST helps to not only detect faint \ion{C}{III}] emission but also to measure it alongside other key UV lines (e.g., HeII, OIII], CIV, [NeV]) that are key for robust diagnostic analyses \citep[e.g.,][]{2019MNRAS.487..333H, 2023MNRAS_Hirschmann, Feltre2016, Gutkin2016, p11_NakajimaMaiolino2022}.
A central challenge that remains, even with JWST's capabilities, is distinguishing the nature of the ionizing sources in \ion{C}{III}] emitters. At high redshifts, classical optical diagnostic diagrams such as the BPT \citep{BaldwinPhillipsTerlevich_1981} and the VO87 \citep{VeilleuxOsterbrock_1987} diagrams become less efficient due to a combination of factors, including different physical conditions like low metallicity \citep[e.g.,][]{2023A&A_Ubler} and the shifting of key optical lines out of observable spectral range of the telescope. Furthermore, applying these diagnostics often requires a high spectral resolution that is not always available for deep, large-scale spectroscopic surveys. Because of this, we investigate the utility and limitations of the OHNO diagram for our sample (see Section~\ref{subsec:OHNO}), noting that recent studies have shown the chosen line ratios are largely degenerate with respect to ionization parameter \citep{2025ApJ_Cleri}.

In this paper, we use JWST data to advance the study of galaxies during the EoR by focusing on the properties of \ion{C}{III}] emitting galaxies. Given the challenges of distinguishing between an AGN and star formation origin at high redshift, a primary goal of our work is to apply both optical and UV diagnostics to identify and distinguish these sources. We perform spectral energy distribution (SED) fitting to estimate physical properties such as stellar mass. Finally, we examine the connection between \ion{C}{III}] emitters and a newly discovered population known as Little Red Dots (LRDs).
\\
The structure of this paper is as follows. Section~\ref{sec:Data_and_sample_selection} describes the observational data and sample selection and outlines our spectroscopic and photometric analysis methodology, including the use of \lime for emission line fitting and \textsc{Bagpipes} for spectral energy distribution modeling. Our main results,as well as our stacking procedure, are presented in Section~\ref{sec:Sample}, with the application of AGN diagnostics detailed in Section~\ref{sec:AGN}. Section~\ref{sec:LRDs} explores the relationship between \ion{C}{III}] emitters and Little Red Dots and Section~\ref{sec:Nature_CIII_emitters} discusses the nature of the \ion{C}{III}] emitters. Finally, we summarize our findings and discuss potential extensions in Section~\ref{sec:Conclusion}.
\\
Throughout this work, we assume a $\Lambda$CDM cosmological model with the following parameters: $H_0 = 70~\mathrm{km~s^{-1}~Mpc^{-1}}$, $\Omega_m = 0.3$, and $\Omega_\Lambda = 0.7$. Magnitudes are expressed in the AB system \citep{Oke1983}, and EWs are reported in rest-frame.

\section{Sample selection and data analysis} \label{sec:Data_and_sample_selection}

\subsection{Spectroscopic data}\label{sec:Spectra_data}

For our analysis, we use \jwst-\NIRSpec\ prism data. The prism, unlike the gratings, provides continuous, wide wavelength coverage (0.6–5.3 $\mu\mathrm{m}$) in a single exposure, which is essential for capturing the full suite of diagnostic lines across our redshift range. The trade-off for this extensive coverage is the low and strongly wavelength-dependent spectral resolution: the resolving power increases nearly linearly from $R\sim30$ at 0.6 $\mu\mathrm{m}$ to $R\sim300$ at 5.3 $\mu\mathrm{m}$. While this resolution limits detailed kinematic measurements and measurements of lines that are too close together, it is suited for detecting and measuring integrated emission line fluxes for AGN diagnostics and population studies, as studied here.

The tail end of the EoR, concluding around 
$z\gtrsim5.3$ \citep[e.g.,][]{Bosman_reionization}, provides a crucial window to study the galaxies responsible for the cosmic transition. The redshift range $5 < z < 7$ targeted in this work is strategically chosen as it probes this final phase of reionization and offers a key observational advantage: it allows the simultaneous observation of both the rest-frame UV emission lines (such as \ion{C}{III}] and \ion{He}{II}) and the optical lines (such as [\ion{O}{III}] and H$\beta$) required for our AGN diagnostic diagrams, as well as H$\alpha$ for investigating broad-line AGN components. We use publicly available \jwst-\NIRSpec-prism data from the CANDELS-Area Prism Epoch of Reionization Survey (CAPERS; GO-6368; PI: Mark Dickinson), an ongoing Cycle 3 JWST Program that is observing up to 10000 high-redshift galaxies split across the Cosmic Evolution Survey (COSMOS), the Ultra-deep Survey (UDS), and the Extended Groth Strip (EGS) fields in the PRISM/CLEAR configuration. In this work, we focus on sources with $5 < z < 7$ from all three fields, finding a total of 549 galaxies.

To complement the CAPERS data\footnote{At the time of the analysis, the reduced CAPERS data from COSMOS and EGS was not in DJA.}, we collect all publicly available \jwst\ spectroscopic observations in the UDS, EGS, COSMOS, Goods North, Goods South and A2744 fields from v3 of the DAWN JWST Archive \citep[DJA,][]{Heintz2024dja, deGraaff2025} up to the 24th of February 2025. We limit our selection to robust (grade=3) spectroscopically identified galaxies at $5 < z < 7$, to match the redshift range of interest from CAPERS. We find a total of 1307 galaxies (corrected for duplicates). 

Combining the data from the DJA sample and the CAPERS sample, we find a total of 1856 unique galaxies. 

%
%\section{Methods} \label{sec:Method}

\subsection{Emission line measurements with \lime}\label{sec:lime}

To identify \ion{C}{III}] emitters in the CAPERS and DJA parent samples, we use the A LIne MEasuring library  \citep[\lime,][]{Fernandez2024Lime}, following a multi-step process. As a first step, we process all galaxies using the spectroscopic redshifts provided by the DJA and CAPERS catalogs. \lime defines rest-frame spectral windows for each emission feature, including the line region and adjacent continuum bands. Emission lines are modeled using Gaussian profiles, each described by three free parameters: amplitude, central wavelength, and velocity dispersion ($\sigma$). For closely spaced emission lines, such as the [\ion{O}{III}] $\lambda\lambda$4959,5007 doublet (hereafter [\ion{O}{III}]), \lime can account for blending by performing a multi component Gaussian fit when specified. This initial analysis yields key measurements including fluxes, equivalent widths, velocity dispersions, and line-based redshifts for various rest-UV and optical emission lines, such as \ion{C}{III}] and [\ion{O}{III}].

We then refine the DJA and CAPERS redshifts using the values from \lime, ensuring the new redshifts did not deviate excessively from the original ones by requiring a difference of less than 0.016 (corresponding to a maximum velocity offset of 600-800 km/s for \ion{C}{III}]). We also verify that the lines used to define the redshift are consistent with the instrumental resolution. This is done by comparing the measured full width at half maximum (FWHM) of the line to the instrumental resolution \footnote{\url{https://jwst-docs.stsci.edu/jwst-near-infrared-spectrograph/nirspec-instrumentation/nirspec-dispersers-and-filters\#gsc.tab=0}} (FWHM$_{\rm inst}$). 

We require a S/N > 2\footnote{We adopt this threshold because we find that \lime tends to slightly underestimate the S/N compared to a direct integration approach. This is due to the fact that in \lime the noise is a combination of the real noise level of the spectrum and the uncertainty of the fit} on the flux of \ion{C}{III}] and we find a sample of 61 \ion{C}{III}] emitters (which will be presented in more detail in Section \ref{sec:Sample}). We then correct the absolute flux calibration of the CIII] emitter spectra for potential slit losses by comparing them to broad-band photometry from \citet{Merlin2024}. Synthetic photometry is generated by integrating each spectrum over standard \NIRCam\ filter bandpasses of F090W, F115W, F150W, F200W, F277W, F356W, and F444W. A single, wavelength-independent correction factor—the weighted average of all available filters—is computed by comparing synthetic and observed fluxes. It is then applied to the entire spectrum and its uncertainties. This process corrects the absolute flux scale without affecting spectral features like equivalent widths or the UV $\beta$ slope. For more details see Appendix \ref{sec:corr_factor}.

\subsection{Measurement of broad $\mathrm{H}\alpha$}
\label{sec:Broad_line_fitting}

We employ a Markov Chain Monte Carlo (MCMC) approach to compare models with and without a broad $\mathrm{H}\alpha$ Gaussian component, and to establish which model better reproduces the observed data. H$\beta$ is not included in this modelling because it is fainter and observed at lower spectral resolution, so we do not expect to detect a broad component there that would not be observed in $\mathrm{H}\alpha$. The first model fits the total observed $\mathrm{H}\alpha$ emission line with a single Gaussian, described by 3 free parameters (peak flux, central wavelength, and FWHM) and a power-law continuum, described by a slope and a normalization coefficient; this results in a total of five free parameters. We set uniform priors for the peak flux, FWHM, and continuum parameters, while we adopt a Gaussian prior for the central wavelength centered on the redshift of the source and with dispersion equal to the redshift uncertainty. The second model adds a second Gaussian component to model the broad emission, characterized by its own amplitude and velocity dispersion, and same central wavelength as the narrow line. This results in a total of seven free parameters: the amplitudes and velocity dispersions for both the narrow and broad components, the common central wavelength, and the two continuum parameters (slope and normalization coefficient). We set analogous priors as for the first model, with the only exception of the prior on the FWHM of the narrow component, $\rm FWHM^{\rm prior}_{narrow line}$. More precisely, $\rm FWHM^{\rm prior}_{\rm narrow~line}$ is set to lie within a continuous interval of 1–2 spectral resolution elements. The size of a spectral resolution element at the central wavelength of the line is computed by accounting for the wavelength-dependent variation of the prism resolution, i.e., converting $R$ to $\rm FWHM$ at a given wavelength using the $R$–$\lambda$ relation from \citet{jakobsen+2022}. At $\mathrm{H}\alpha$ wavelength, the prism spectral resolution corresponds to a velocity width of $\sim 1000$--$1400~\mathrm{km\,s^{-1}}$ across our redshift range ($5 < z < 7$). The best fitting values and associated errors are derived from the 50th, 16th and 84th percentiles of the posterior distributions, respectively.

Our primary diagnostic is the Bayesian Information Criterion (BIC), calculated from the $\chi^2$ values of each fit. $\Delta \mathrm{BIC}=\mathrm{BIC_{ narrow}} - \mathrm{BIC_{broad}}$ > 6  indicates strong evidence for a broad component, while $\Delta \mathrm{BIC}$ > 2 suggests tentative evidence \citep[e.g.,][]{juodžbalis2025jadescomprehensivecensusbroadline}. As a secondary check, we compare the reduced $\chi^2$ values, favoring the model with the lower value. Finally, we visually inspect the residuals and line profiles to ensure consistency. 

\subsection{Physical properties with \textsc{Bagpipes}} \label{sec:Muv}

To estimate the physical properties of a galaxy, we use the Bayesian Analysis of Galaxies for Physical Inference and Parameter EStimation \citep[\textsc{Bagpipes},][]{Carnall_2018} code to fit photometric data. Our analysis employs the photometric catalog presented by \citet{Merlin2024}, comprising 16 bands from \textit{HST} and \textit{JWST}, drawn from large public surveys, including CANDELS \citep{Grogin2011,Koekemoer2011} and CEERS \citep{Bagley2023,Finkelstein2025}, that provide continuous wavelength coverage from $0.44$ to $4.44~\mu\mathrm{m}$. The filter set includes: \textit{HST}/ACS optical bands (F435W, F606W, F775W, F814W); \textit{HST}/WFC3 near-infrared bands (F105W, F125W, F140W, F160W); and \textit{JWST}/NIRCam bands (F090W, F115W, F150W, F200W, F277W, F356W, F410M, F444W). These data cover the same fields from which our spectroscopic sample was selected. This coverage enables robust spectral energy distribution fitting across the rest-frame UV to optical for 55 out of 61 of our $z = 5-7$ sample. The catalog provides total flux measurements and associated uncertainties in microJanskys ($\mu$Jy).

We perform SED fitting with \textsc{Bagpipes} assuming  a delayed-$\tau$ star formation history, characterized by two main parameters: the galaxy age (varied from 0.001 Gyr up to the Universe’s age at the observed redshift) and the $\tau$ parameter (ranging from 0.1 to 10 Gyr), which controls the timescale of star formation. The total stellar mass formed is allowed to vary between $10^1$ and $10^{15}\ M_\odot$, and the metallicity is explored within 0 to 0.5 times the solar value. Dust attenuation is modeled using a Calzetti law \citep[e.g.,][]{Calzetti_2000_DustAttenuation} with $A_V$ ranging from 0 to 2 magnitudes \citep[e.g.,][]{2025Nat_Markov}. Additionally, nebular emission is included, with the ionization parameter $\log U$ allowed to vary between –4 and 0. By fitting this model to the observed data, \textsc{Bagpipes} infers the best-fitting set of physical parameters that describe the galaxy’s evolution and current state.

To account for potential AGN activity, we expand the base model adding the AGN model described in \citet{Carnall2023}, which includes a continuum component, modeled as a broken power law with a break at 5100 \r{A}, and broad $\mathrm{H}\alpha$ and $\mathrm{H}\beta$ emission. Free parameters are the $\mathrm{H}\alpha$ flux and line width, the continuum flux at 5100 \r{A}, and the power-law slopes. The broad $\mathrm{H}\beta$ flux is derived from $\mathrm{H}\alpha$ considering Case~B recombination and the same line width. The ranges for the stellar+nebular components remain the same.

The UV absolute magnitude is estimated for each galaxy from the best-fitting SED template using a top-hat filter centered on $\lambda_{\rm rest}$ =1500 \r{A} with a width of 100 \r{A}. This is then converted to an absolute magnitude using
\begin{equation}
M_{\rm UV}=m_{1500}-5(\log(D_L)-1)+ 2.5\log(1 + z)
\end{equation}
where $m_{1500}$ is the apparent magnitude at $\lambda_{\rm rest}$ =1500 \r{A} and $D_L$ is the luminosity distance in parsecs.
The uncertainty on $M_{\text{UV}}$ is estimated by simple error propagation on the flux following e.g. \citet{2025A&A_Llerena, 2025ApJ_Pahl}. 

We derive H$\alpha$-based star-formation rates (SFRs) estimating the H$\alpha$ emission line flux, applying dust corrections using the $A_V$ values estimated by \textsc{Bagpipes} SED fitting, converting to H$\alpha$ luminosity $L_{\mathrm{H}\alpha}$, and finally converting $L_{\mathrm{H}\alpha}$ to SFR using the calibration from \cite{Reddy2022} as \begin{equation}
    \mathrm{SFR} = L_{\mathrm{H}\alpha} \times 10^{-41.67} \; M_\odot \, \mathrm{yr}^{-1}.
\end{equation}

Finally, we use \textsc{Bagpipes} best-fitting models to derive the stellar mass of our sources. For each galaxy, we compare the reduced chi-squared ($\chi^{2}{\mathrm{red}}$) values derived for fits performed with and without an AGN component, and we consider the stellar mass resulting from the fit with the lowest $\chi^{2}{\mathrm{red}}$ (see Sect. \ref{sec:AGN_comp} for details).

Specifically, for CEERS-1324, since no photometry is available, we perform a spectroscopic fit with \textsc{Bagpipes}. We add a second order polynomial with Gaussian priors on its coefficients to correct for potential flux calibration mismatches between the model and the observed spectrum, and a white noise scaling model to represent wavelength-independent uncertainties. We also include the instrumental resolution curve extracted from the JWST/NIRSpec prism dispersion file to ensure the spectral model is convolved to the appropriate resolving power over the fitted wavelength range. Additionally, in order to promote fast convergence, we fix $\log U$ to a value of $-2$, which is close to the average ionization parameter found from the photometric SED fitting of all the other sources in our sample.

%The procedure for selecting the best-fitting model and determining the final value is described in detail in Appendix \ref{sec:model_selection}. 

\section{A sample of \ion{C}{III}] emitters}\label{sec:Sample}

\begin{figure}[t!]
    \centering
    \begin{subfigure}[b]{0.49\textwidth}
        \centering
        \includegraphics[width=\textwidth]{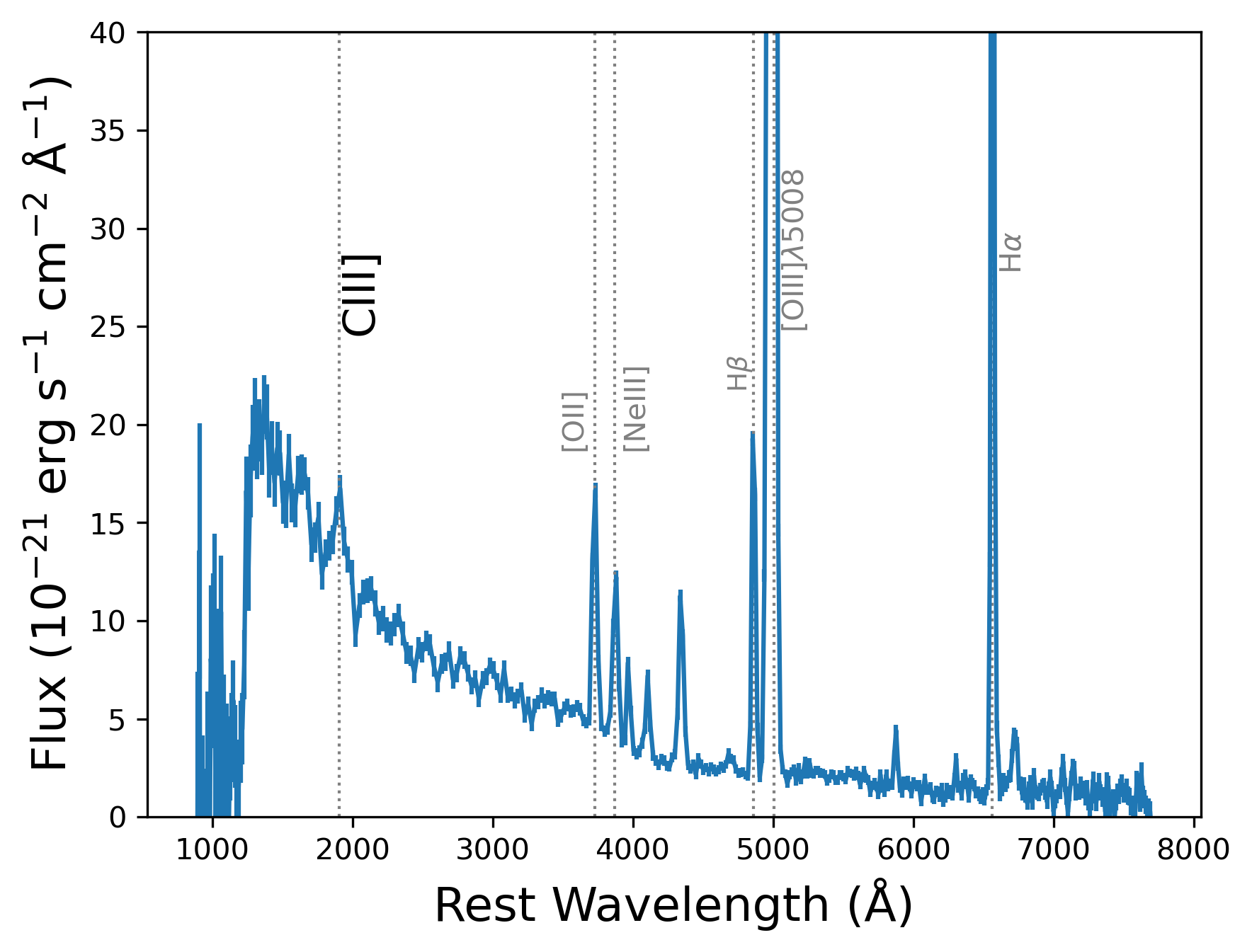}
        \caption{COSMOS-67506 (z = 5.8666) from the COSMOS field with an EW(\ion{C}{III}]) = 21.85 $\pm$ 8.25~\r{A}. }
        \label{fig:galaxy_cosmos_example}
    \end{subfigure}
    \hfill
    \begin{subfigure}[b]{0.49\textwidth}
        \centering
        \includegraphics[width=\textwidth]{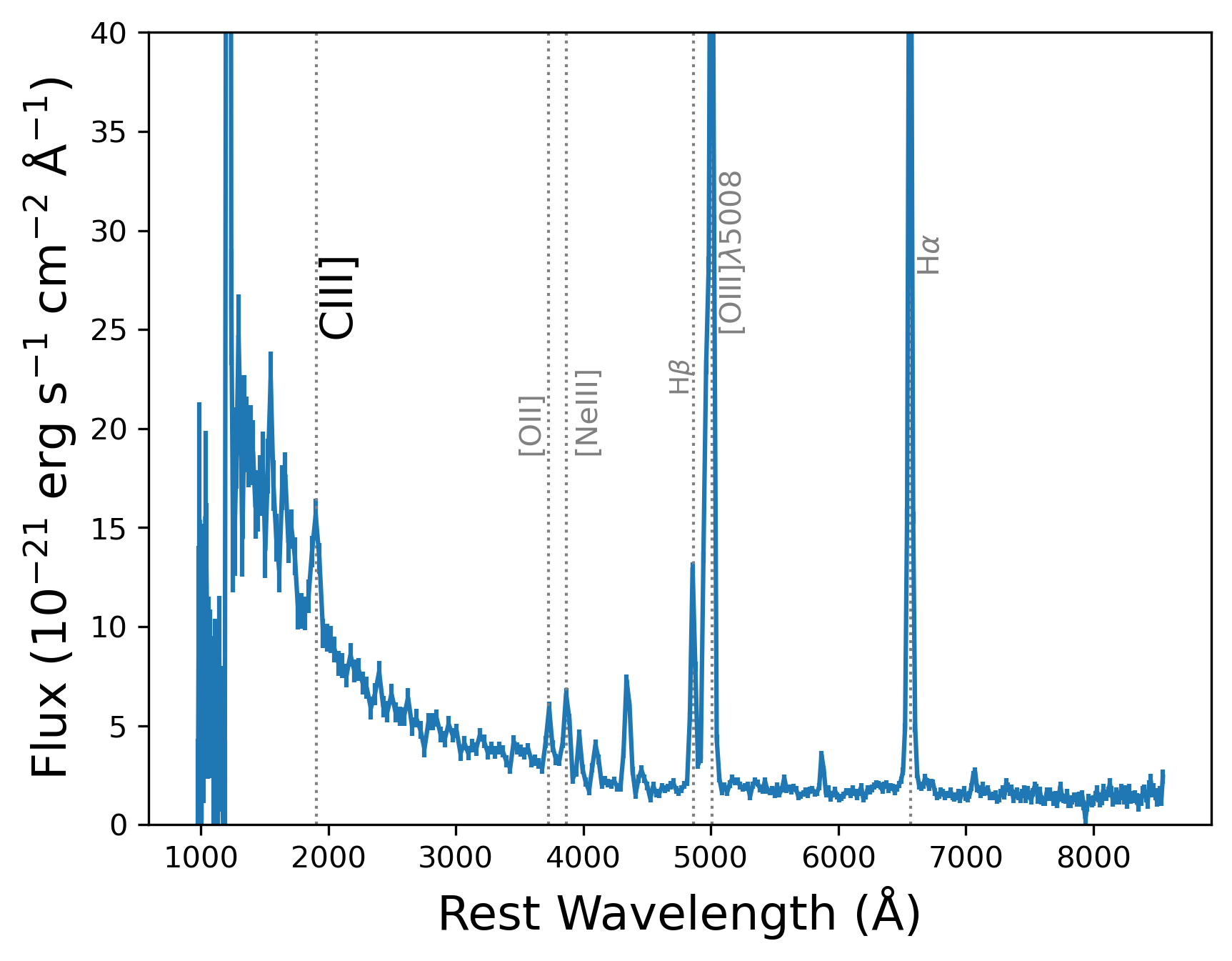}
        \caption{CAPERS-9244 (z = 5.203) from the EGS field with an EW(\ion{C}{III}]) = 24.24 $\pm$ 9.28~\r{A}.}
        \label{fig:galaxy_egs_example}
    \end{subfigure}
    \caption{Rest-frame spectra of \ion{C}{III}]-emitters from different surveys. Both panels show the rest-frame wavelength in Ångströms on the x-axis and flux density in units of $10^{-21}$ erg s$^{-1}$ cm$^{-2}$ Å$^{-1}$ on the y-axis.}
    \label{fig:galaxy_examples_combined}
\end{figure}

We identify 61 \ion{C}{III}] emitters out of 1896 galaxies analysed with \lime: two examples of our spectra are shown in Fig. \ref{fig:galaxy_examples_combined}.

\begin{figure}[t!]
    \centering    \includegraphics[width=\columnwidth]{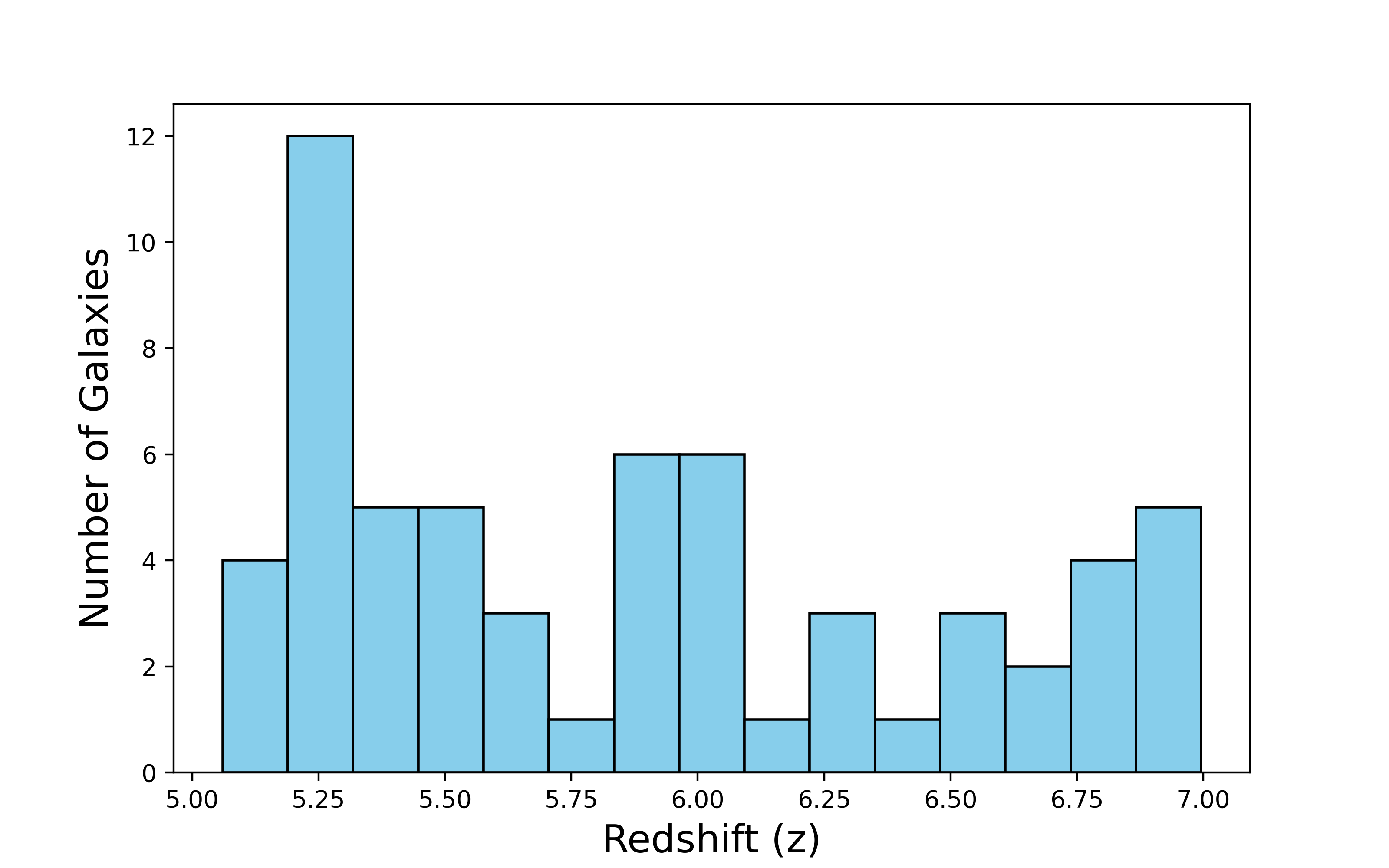}
    \caption{Redshift distribution of our galaxy sample. The histogram shows the number of galaxies as a function of redshift between 5 and 7, with a pronounced peak at 5.25.}
    \label{fig:redshift_distribution}
\end{figure}

Our sample comprises galaxies spanning redshifts from 5 to 7, with the distribution peaking around $z=5.25$ (as can be seen in Figure \ref{fig:redshift_distribution}). The sample includes 24 galaxies with z > 6 and 37 galaxies with z $\leq$ 6, resulting in a median redshift of 5.87.

\begin{figure}[t!]
    \centering    \includegraphics[width=\columnwidth]{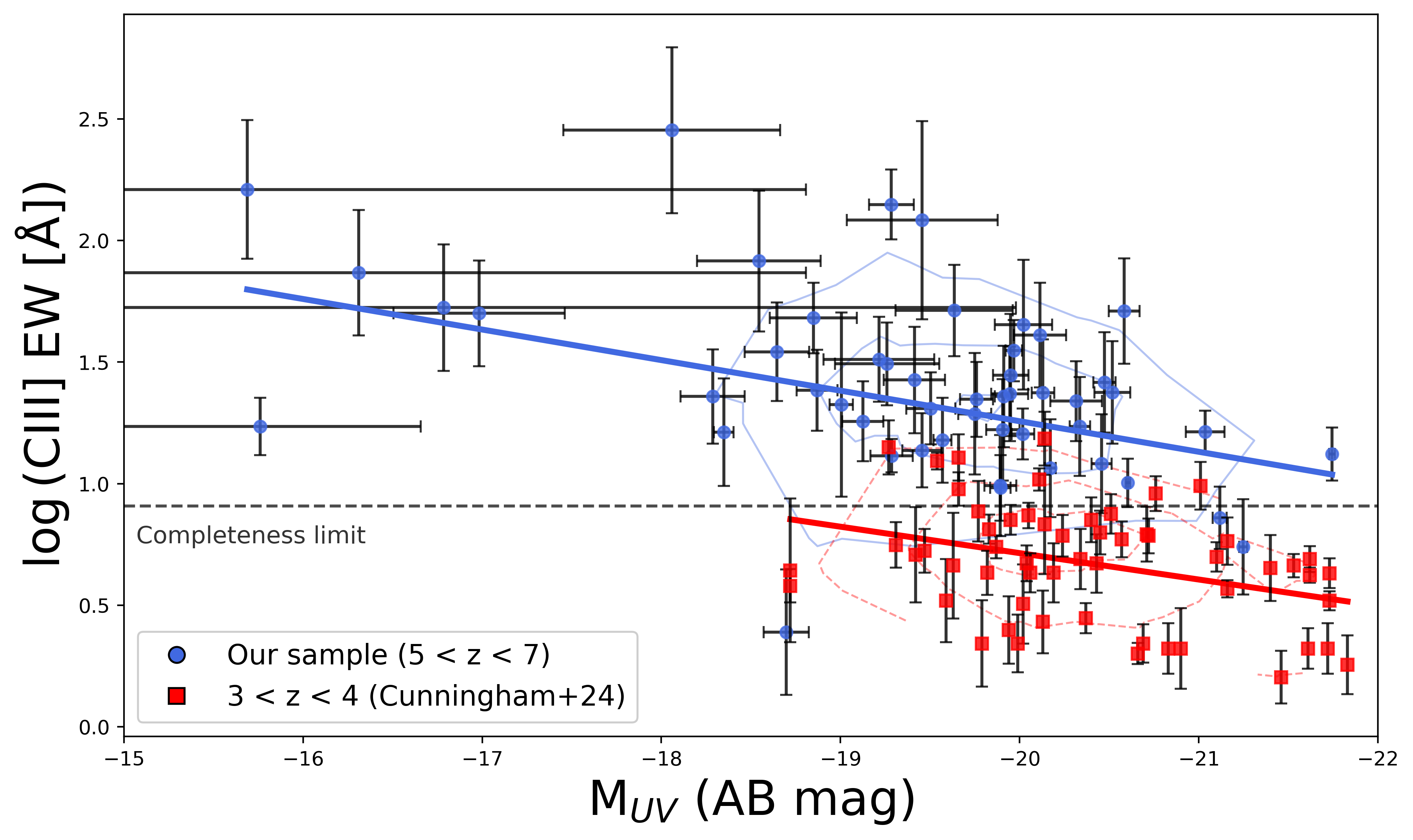}
    \caption{Rest-frame \ion{C}{III}] equivalent width (EW; in \AA) versus absolute UV magnitude ($M_{\text{UV}}$; AB mag). The blue points correspond to our sample of \ion{C}{III}] emitters with redshifts between 5 and 7. The red points correspond to a VANDELS sample compiled by \cite{Cunningham_2024} with redshifts between 3 and 4. Contour plots and linear regression fits are also shown.}
    \label{fig:EW_CIII_vs_Muv}
\end{figure}

The EW(\ion{C}{III}]) in our sample span nearly two orders of magnitude, from 2.5 ~\AA\ to 284.2 ~\AA. The distribution has median EW of 22.8~\AA\ (16th--84th percentile: 12.5--51.5~\AA) and a mean of 39.2~\AA. We compare the rest-frame EW(\ion{C}{III}])  of our $5 < z < 7$ galaxies against lower-redshift ($3 < z < 4$) emitters from the VANDELS survey \citep{Cunningham_2024}. At fixed $M_{\text{UV}}$, our sample exhibits systematically stronger EW(\ion{C}{III}]) compared to their lower-redshift analogs (see Figure \ref{fig:EW_CIII_vs_Muv}). Quantitatively, the median EW for the combined $3 < z < 4$ sample is 4.7~\AA\, compared to our 22.8~\AA. This corresponds to an increase of $\sim$0.67~dex in \ion{C}{III}] EW from $z = 3-4$ to $z = 5-7$, indicating strong redshift evolution in the line strength.

This enhancement could stem from several factors. Observational biases may play a role, as JWST/NIRSpec prism spectroscopy lacks the sensitivity to detect \ion{C}{III}] emission with low equivalent width (EW $\leq 5$–$10~\AA$) at these redshifts, thus potentially incompletely sampling the low-EW population. In fact, 
the EW(\ion{C}{III}]) measured from the stacked spectrum of non-detections in our sample (see Table \ref{tab:stack_all_lines}) is $5.7 \pm 0.3$~\AA, consistent with the typical values found at lower redshifts. This suggests that we are likely missing the low-EW population due to observational limitations. Indeed, from the completeness analysis of our sample, we estimate a detection limit of 8.1~\AA, below which the number counts drop to 50\% of the maximum frequency. This selection bias means we are incompletely sampling the low-EW population at $z = 5-7$.

Nevertheless, the absence of similarly high EW values at lower redshifts might indicate a genuine evolutionary trend. In other words, the systematically elevated EWs in our high-redshift sample might indicate that galaxies at $z = 5-7$ inherently produce stronger \ion{C}{III}] emission.

This evolutionary trend aligns with results from \cite{robertsborsani2024extremesjwstspectroscopicbenchmark}, and may reflect harder radiation fields, lower metallicities, and/or higher ionization parameters in the early universe \citep[e.g.,][]{Llerena2022,p12_nakajima2018,Nakajima2023, 2023ApJ_Trump,2023ApJ_Sanders,2024ApJ_Sanders,Curti2023,2023A&A_Cameron}.

On the other hand, the different distribution of EW(\ion{C}{III}]) might be due to the fact that the JWST sample is actually dominated by AGN rather than star forming galaxies. Indeed, JWST has revealed the existence of a much larger population of high redshift AGN than expected \citep[e.g.,][]{juodžbalis2025jadescomprehensivecensusbroadline,2025A&A_Scholtz,2025ApJ_Treiber}. 
Our analysis of the \ion{C}{III}] properties will help us discriminate between the two possibilities. 

Additionally, we find a significant correlation between EW(\ion{C}{III}]) and $M_{\text{UV}}$ in our sample, with a slope of $0.126 \pm 0.039$. This indicates that fainter galaxies (higher $M_{\text{UV}}$ values) tend to have stronger \ion{C}{III}] emission relative to their continuum, consistent with the behavior observed in lower-redshift samples. The \cite{Cunningham_2024} sample also reveals a positive correlation slope of $0.109 \pm 0.039$ in their data.

\begin{figure}[t!]
    \centering    \includegraphics[width=\columnwidth]{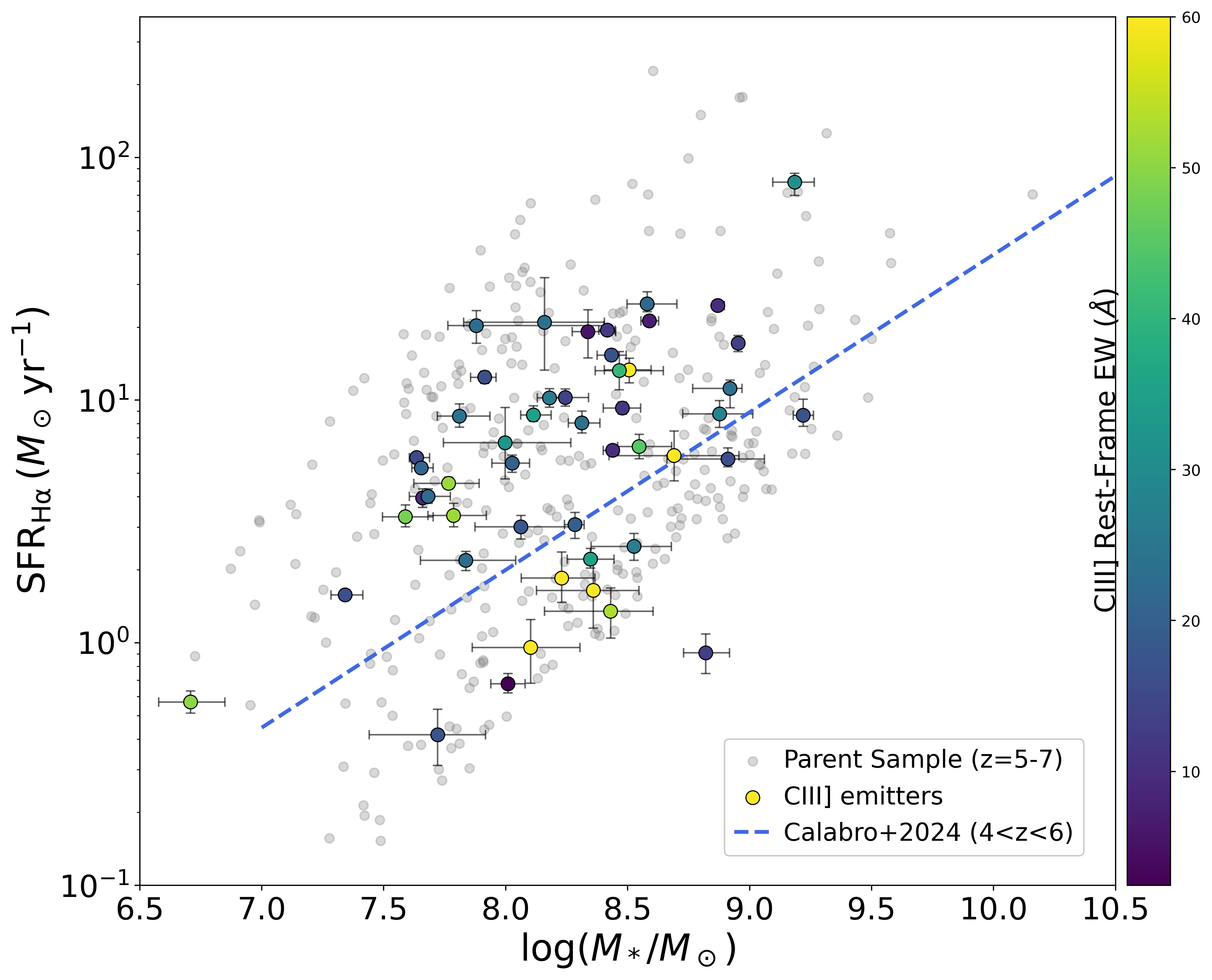}
    \caption{Plot of the SFR vs stellar mass of our sample compared to the star formation of field galaxies at z=5-7. Our sample is color coded by the rest frame \ion{C}{III}] EW value. The general parent population in the same redshift range is plotted as grey circles. The blue line represents the relationship found by \cite{Calabr__2024} and derived with SFR based on $\mathrm{H}\alpha$ for a similar spectroscopic sample.}
    \label{fig:Main_sequence_diagram}
\end{figure}

In Fig. \ref{fig:Main_sequence_diagram}, we show the star formation mass sequence diagram \citep[SFR vs stellar mass,][]{2014ApJS..214...15S} to check where our \ion{C}{III}] emitters lie with respect to the general population. We color code our galaxies by the EW(\ion{C}{III}]) value. We do not include the \ion{C}{III}] emitters for which we do not have photometry (since we are not able to perform the SED fitting) and we also do not include our sample of Little Red Dots (as explained in Section \ref{sec:LRDs}). In the background, we plot a reference sample of spectroscopic galaxies from the same surveys with $5<z<7$. This represents the general parent population. We also plot the main sequence line from \cite{Calabr__2024} derived with $\mathrm{H}\alpha$ based SFRs for a similar spectroscopic sample. As shown in Figure \ref{fig:Main_sequence_diagram}, our \ion{C}{III}] emitters are distributed across the entire main sequence, showing no strong systematic trend. However, we note that galaxies lying below the \cite{Calabr__2024} line are more likely to exhibit strong equivalent widths, although such emitters are not confined exclusively to this region.

\subsection{Stacked spectra}\label{sec:stacking_method}
To characterize the global spectral features of the \ion{C}{III}] emitters in the sample, we perform a stacking analysis, binning the sample by EW(\ion{C}{III}]). This allows us to increase the S/N of the final spectra. For this purpose, we separate the galaxies into two EW(\ion{C}{III}]) bins. Since the median EW in the sample is 23~\AA, we perform a stack of weak emitters (EW$\leq$22.8 \r{A}) and strong emitters (EW$>$22.8 \r{A}). Each bin is populated by 31 and 30 galaxies, respectively. 

\begin{figure}[t!]
    \centering
    \includegraphics[width=\columnwidth]{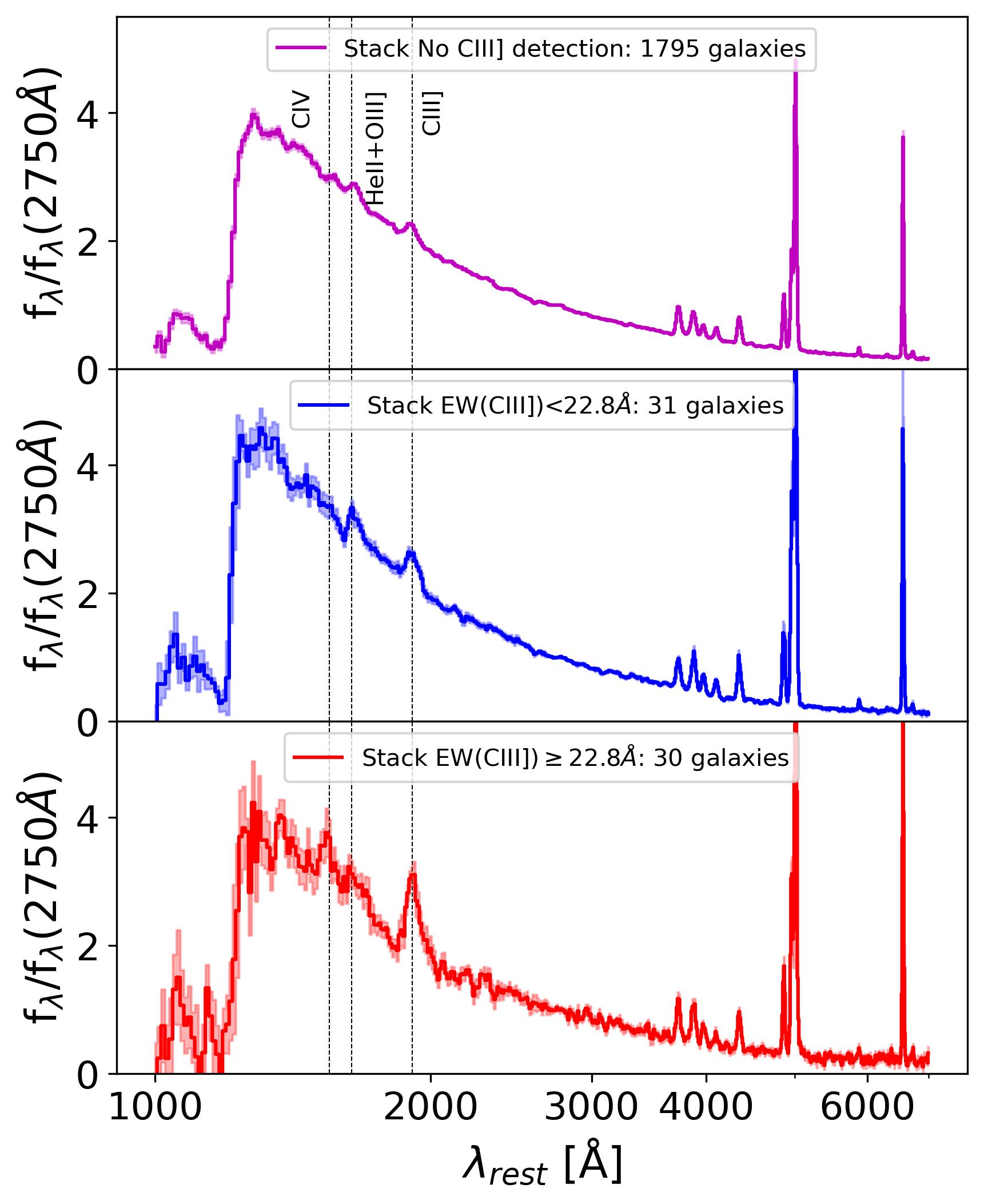}
    \caption{Stacked spectra for the sample of non-\ion{C}{III}] emitters (top panel), weak (middle panel), and strong (bottom panel) \ion{C}{III}] emitters. In all panels, the shaded regions represent the 1$\sigma$ uncertainty of the stacking. The dashed black vertical line indicates the position of rest-frame UV emission lines.}
    \label{fig:plot_stacks}
\end{figure}

The stacking is based on a non-weighted scheme. All the individual spectra in the sample are first shifted into the rest-frame using the systemic redshift based on [OIII]$\lambda$5007\r{A} and then they are resampled onto a common grid. To define a common wavelength grid for the stacking, we first compute the minimum spectral sampling ($\Delta\lambda$) of each individual spectrum in the observed frame $>1400~(1+z)~\rm \AA$, which corresponds to 70.54~\r{A}. We then divide by (1+$z_{\rm mean}$) as the characteristic rest-frame wavelength spacing, where $z_{\rm mean}$ is the mean redshift of the stacked galaxies in each bin. This sampling provides a representative dispersion for the sample and ensures a consistent grid when resampling spectra obtained with the NIRSpec/prism, whose wavelength dispersion varies with wavelength and redshift. The mean systemic redshift of the sample is $z_{\rm mean}\sim6.06$ for the strong emitters and $z_{\rm mean}\sim5.71$ for the weak emitters. After that, the spectra are normalized to the median flux between 2700 and 2800~\r{A}, a wavelength range that is free of bright emission lines in the individual spectra. The final flux at each wavelength is taken as the median of all the individual flux densities. The 1-$\sigma$ error spectrum is estimated by a bootstrap re-sampling of the individual flux densities for each wavelength and taking the standard deviation of the resulting median stacked spectra. The stacked spectra are shown in Fig. \ref{fig:plot_stacks}.

We also stack the spectra of non-\ion{C}{III}] emitters (those galaxies from our parent sample that had a S/N < 2 on the flux of \ion{C}{III}]), which includes 1795 galaxies with a mean redshift $z_{\rm mean}=5.8$, after excluding overlapping sources in the multiple surveys. We follow the same methodology as for \ion{C}{III}] emitters and the resulting stacked spectrum is shown on the top panel in Fig. \ref{fig:plot_stacks}.

Similarly to the methodology used for individual galaxies, we use \lime \citep{Fernandez2024Lime} to measure the fluxes and EWs of the UV and optical emission lines. For the considered emission lines, we assume a single Gaussian model.

\section{AGN diagnostics}\label{sec:AGN}

Understanding whether \ion{C}{III}] emission in galaxies is powered by star formation or AGN is crucial for interpreting their physical conditions. AGN can significantly enhance high-ionization emission lines, including \ion{C}{III}], because of their hard ionizing spectra, while SFG produce these lines through stellar processes such as massive young stars and shocks. Without properly distinguishing AGN from SFG, studies of \ion{C}{III}] emitters risk misattributing the source of ionization, leading to incorrect estimations of stellar masses and SFRs. Therefore, robust AGN diagnostics are essential. 

\subsection{Broad emission line and BH mass estimate}

Broad $\mathrm{H}\alpha$ emission is a classic signature of an AGN (where the broad-line region is unobscured), originating from high-velocity gas in the vicinity of the central supermassive black hole \citep[e.g.,][]{2001AJ_Berk, 1969Nature_Lynden-Bell}. The presence of a broad component—typically with a FWHM of several thousand km/s—indicates gas motions governed by the gravitational potential of the black hole, providing strong evidence for AGN activity distinct from star formation. Detecting such broad lines, even in low-resolution spectra, provides strong evidence for AGN activity and is a key diagnostic in distinguishing AGNs from SFGs.

Detecting broad emission is particularly challenging with the NIRSpec prism as a result of its lower spectral resolution which can blur broad and narrow features together. Despite this limitation, our three-tier approach (see Section \ref{sec:Broad_line_fitting}) allow us to robustly identify broad $\mathrm{H}\alpha$ emission in 15 out of 60 galaxies, 5 of them being LRDs (see Section \ref{sec:LRDs}).

%Include here the explanation of the embedded [NII] line.
When assessing the presence of a broad H$\alpha$ component, we additionally verify that the apparent line broadening was not driven by contamination from the merged \ion{N}{II}] emission. Among the 15 sources exhibiting a broad H$\alpha$ component, three have FWHM values below 2000 km s$^{-1}$, while the remaining sources show FWHM $>2000$ km s$^{-1}$, which cannot be reproduced by \ion{N}{II}] emission alone. For the three lower-FWHM sources, we performe an additional fit including only a narrow H$\alpha$ component and the \ion{N}{II}] doublet, and compare this model to one including both narrow and broad H$\alpha$ components following the same procedure as previously. We find that for only galaxy GDS-202208, the model including \ion{N}{II}] alone is preferred over the one with a broad H$\alpha$ component. This source is therefore excluded from the sample of broad H$\alpha$ galaxies.

\begin{figure}[t!]
    \centering
    \includegraphics[width=\columnwidth]{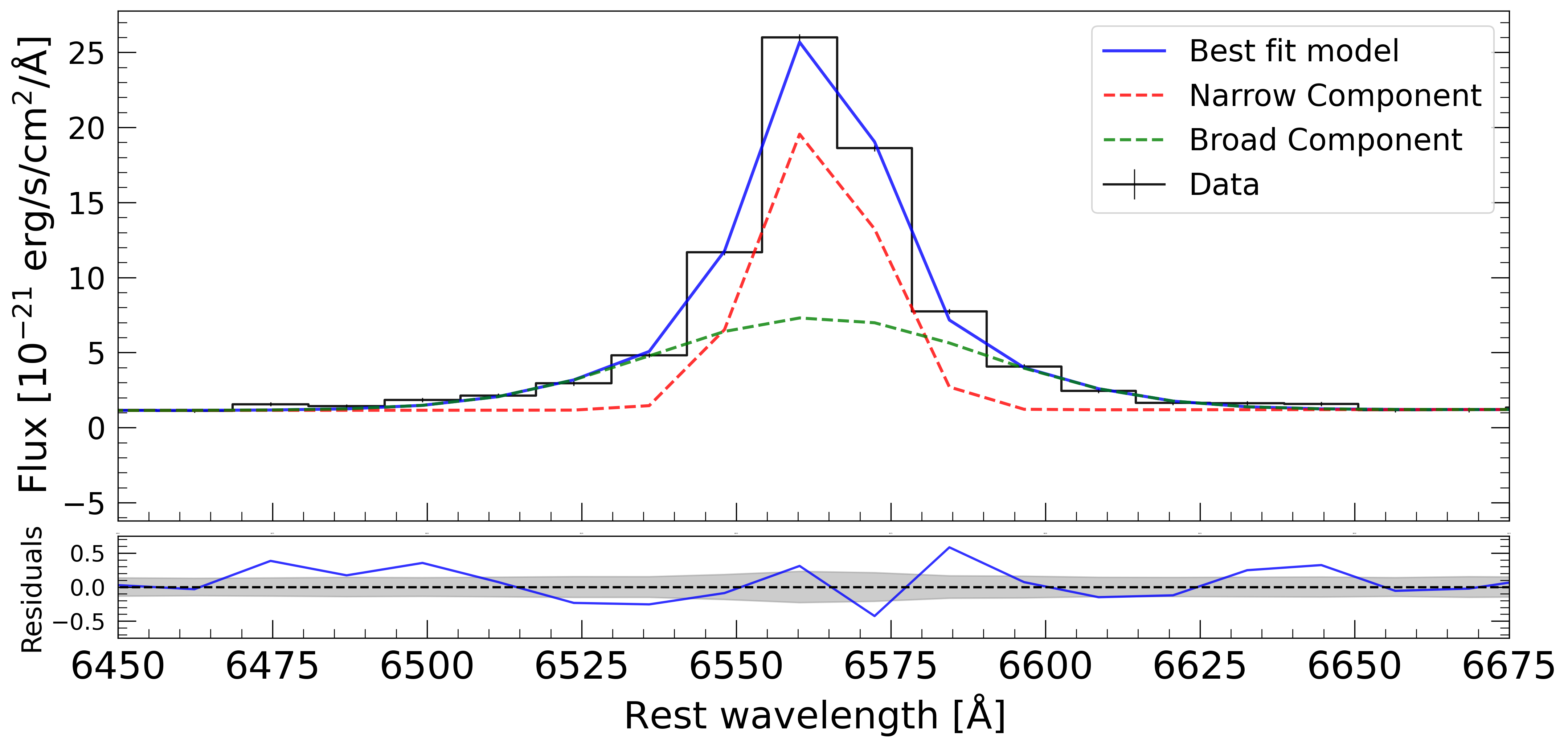}
    \caption{Example of a galaxy with a significant broad H$\alpha$ component: GDS-13704 at $z = 5.93$. The black line shows the observed spectrum. The blue line represents the best-fit model using a double Gaussian profile. The red line shows the narrow component, while the green line corresponds to the broad component. The x-axis shows the rest-frame wavelength in Angstroms, and the y-axis shows the flux density in units of $10^{-21}$ erg s$^{-1}$ cm$^{-2}$\AA$^{-1}$. This source has a $\Delta \mathrm{BIC} = 34.5$, providing strong evidence for the presence of broad-line emission.}
    \label{fig:broad_line_example}
\end{figure}

An example of a galaxy exhibiting a broad H$\alpha$ component is shown in Figure \ref{fig:broad_line_example}, which presents the source GDS-13704 from the GOODS-South field at a $z=5.93$. This source has a $\Delta \mathrm{BIC} = 34.5$, indicating strong evidence for the presence of a broad component.

Based on the detected broad $\mathrm{H}\alpha$ emission, we derive BH masses for our 14 galaxies adopting the relation from Equation 1 in \citet{reines2015}. From the BH mass measurements, we calculate the Eddington luminosity as
$L_{\rm Edd} = 1.3\times 10^{38} ~(M_{\rm BH, \textrm{H}\alpha}/M_\odot) {~\rm erg ~s^{-1}}$. We also derive the bolometric luminosity ($L_{\rm bol}$) of the AGN using the continuum luminosity at 3000 \AA ~and the bolometric correction presented by \citet{richards2009}. From the $L_{\rm Edd}$ and $L_{\rm bol}$, we infer the corresponding Eddington ratios $\lambda_{\rm Edd}=L_{\rm bol}/L_{\rm Edd}$. Results are reported in Table \ref{tab:bh-mass}.

\begin{figure}
    \centering
    \includegraphics[width=0.9\linewidth]{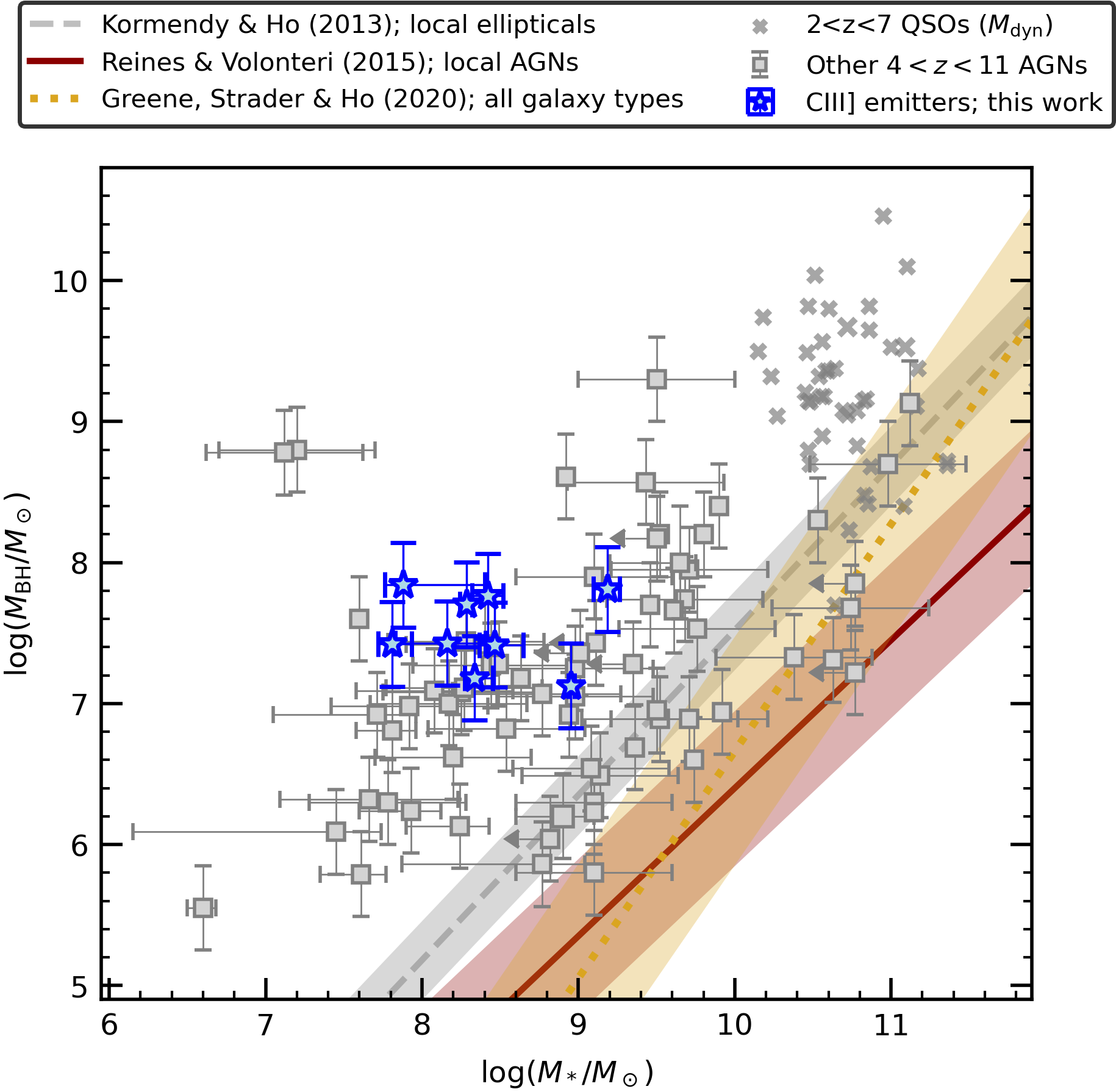}
    \caption{{\bf Black hole mass versus stellar mass.} The results for our 9 \ion{C}{III}] emitters with broad $\mathrm{H}\alpha$ emission are compared with $4<z<10$ AGNs \citep[gray squares,][full references in the text]{Harikane2023b,kocevski2023,juodžbalis2025jadescomprehensivecensusbroadline,wang2024,ubler2024,akins2025,kokorev2023,larson2023,bogdan2024,Maiolino2024}, with a compilation of $2<z<7$ QSOs \citep[][and references therein]{tripodi2024a}, and with local scaling relations \citep[dashed grey, solid red and dotted yellow lines, respectively from refs.][]{kormendy2013,reines2015,greene2020}.}
    \label{fig:mbh-mstar}
\end{figure}

Figure \ref{fig:mbh-mstar} shows our sample of \ion{C}{III}] emitters in the $M_{\rm BH}-M_{*}$ diagram compared with other known AGNs at $4<z<11$ \citep{juodžbalis2025jadescomprehensivecensusbroadline,2023A&A_Ubler,maiolino2025b,marshall2025,parlanti2024,rinaldi2025,ubler2025,fei2025,bogdan2024,larson2023,akins2025,Harikane2023b,kocevski2023,kokorev2024b,tripodi2025,ding2023,Harikane2023b,wang2024,ubler2024,Maiolino2024}, and a compilation of QSOs at $2<z<7$ \citep[][and references therein]{tripodi2024a}\footnote{For QSO hosts we consider the dynamical mass as a proxy for the stellar mass, since the stellar mass is not yet available for these sources.}. We exclude from the plot our 5 LRDs given the significant uncertainty in deriving their stellar mass (see Section \ref{sec:LRDs}). Similarly to the majority high-z AGNs, our sample of \ion{C}{III}] emitters lies above the local relations \citep{kormendy2013,reines2015,greene2020}, indicating a more rapid evolution of BHs with respect to their hosts in this early phase of their evolution. 
We additionally check for a possible relation between the ratio of $\log(M_{\rm BH,\mathrm{H}\alpha}/{\rm M_{\rm *}})$ and the EW(\ion{C}{III}]) and we find no significant correlation.

\begin{table*}[]
    \centering
    \begin{tabular}{c|cccccc}
    \hline
    \hline 
       ID  & $L_{\mathrm{H}\alpha, \rm BLR}$ & $\log(M_{\rm BH}/\rm M_\odot)_{\mathrm{H}\alpha}$ & $\log(L_{\rm Edd}/\rm erg~s^{-1})$ & $\log(L_{\rm bol}/\rm erg~s^{-1})$ & $\lambda_{\rm Edd}$ & $\log(M_{\rm *}/\rm M_\odot)$ \\
           & [$\rm 10^{42} ~erg ~s^{-1}$] & [dex] & [dex] & [dex] & & [dex]\\
           \hline
         CEERS-397 & $3.31_{-1.03}^{+2.20}$ & 7.18 & 45.29 & $43.67 \pm 0.05$ & 0.024 & $8.3 \pm 0.1$\\[0.1cm]
JADES-38147 & $7.17_{-0.66}^{+0.63}$ & 7.84 & 45.96 & $43.0 \pm 0.2$ & 0.001 & $7.9_{-0.1}^{+0.5}$\\[0.1cm]
JADES-167639 & $0.95_{-0.14}^{+0.14}$ & 7.70 & 45.82 & $43.1 \pm 0.1$ & 0.002 & $8.28 \pm 0.04$\\[0.1cm]
RUBIES-16915 & $1.44_{-0.42}^{+0.48}$ & 7.13 & 45.24 & $43.85 \pm 0.04$ & 0.04 & $8.95 \pm 0.02$\\[0.1cm]
RUBIES-174752 & $1.67_{-0.35}^{+0.48}$ & 7.43 & 45.54 & $43.2 \pm 0.1$ & 0.005 & $8.2_{-0.3}^{+0.2}$\\[0.1cm]
RUBIES-37427 & $2.00_{-0.43}^{+0.38}$ & 7.42 & 45.53 & $43.2 \pm 0.2$ & 0.004 & $8.5_{-0.1}^{+0.2}$\\[0.1cm]
CAPERS-52661 & $8.78_{-0.52}^{+0.53}$ & 7.81 & 45.92 & $43.2 \pm 0.1$ & 0.002 & $9.2 \pm 0.1$\\[0.1cm]
CAPERS-9244 & $0.93_{-0.15}^{+0.20}$ & 7.42 & 45.54 & $42.91 \pm 0.05$ & 0.002 & $7.8 \pm 0.1$\\[0.1cm]
CEERS-1334 & $4.07_{-0.36}^{+0.38}$ & 7.76 & 45.88 & $43.2 \pm 0.1$ & 0.002 & $8.42 \pm 0.1$\\[0.1cm]
COSMOS-61234 & $11.45_{-0.24}^{+0.23}$ & 8.03 & 46.15 & $43.27 \pm 0.05$ & 0.001 & N/A\\[0.1cm]
GDS-13704 & $1.14_{-0.06}^{+0.06}$ & 7.46 & 45.57 & $42.6 \pm 0.1$ & 0.001 & N/A\\[0.1cm]
RUBIES-49140 & $59.24_{-0.71}^{+0.75}$ & 8.50 & 46.62 & $43.1 \pm 0.1$ & 0.0003 & N/A\\[0.1cm]
UNCOVER-41225 & $2.18_{-0.23}^{+0.23}$ & 7.51 & 45.63 & $43.2 \pm 0.1$ & 0.003 & N/A\\[0.1cm]
CAPERS-35805 & $10.53_{-0.30}^{+0.30}$ & 8.13 & 46.25 & $43.2 \pm 0.1$ & 0.001 & N/A\\[0.1cm]
         \hline
         \hline
    \end{tabular}
    \caption{$\mathrm{H}\alpha$ luminosity, BH mass, Eddington luminosity, bolometric luminosity, Eddington ratio, and stellar mass for the 14 galaxies exhibiting broad $\mathrm{H}\alpha$ emission. We consider an usual 0.3 dex uncertainty for BH masses and Eddington luminosities, accounting for calibration systematics.}
    \label{tab:bh-mass}
\end{table*}

%\subsection{MEx}
\subsection{UV diagnostics}

Other reliable AGN diagnostics are those based on UV lines. Specifically, diagnostics involving the \mbox{\ion{He}{II} $\lambda$1640} and [\ion{Ne}{V}] lines show an effective discriminating power \citep[e.g.,][]{Feltre2016, Gutkin2016}. Unfortunately, the lines needed to employ those diagnostics are not easily detected individually. In the case of \mbox{\ion{He}{II} $\lambda$1640}, the prism resolution is not enough to separate it from \mbox{\ion{O}{III}] $\lambda$1665} in most cases, especially when the SN is low.

In our case, we aim to employ two AGN diagnostics based on the \mbox{\ion{He}{II} $\lambda$1640} emission line, i.e. the EW(\ion{C}{III}])  vs \mbox{\ion{C}{III}]/\ion{He}{II} $\lambda$1640} and the EW(\ion{C}{IV}]) vs  \mbox{\ion{C}{IV}/\ion{He}{II} $\lambda$1640} \citep{p12_nakajima2018}. To constrain the \mbox{\ion{He}{II} $\lambda$1640} flux, we measure the merged flux of the (\ion{He}{II} + \ion{O}{III}]) feature and subsequently estimated the \ion{He}{II} contribution alone, based on the \ion{O}{III}]/\ion{He}{II} ratio measured in the stacked spectra, as described in Appendix \ref{sec:ratio_calibration}. In particular, we use the derived \ion{O}{III}]/\ion{He}{II} ratio of 0.87$\pm$0.38. Note that this value is similar to the ratios previously found in both star-forming and AGN host galaxies at intermediate redshift  \citep[e.g.,][]{Amorin2017, Llerena2022}.

\begin{figure}[t!]
    \centering
    \includegraphics[width=\columnwidth]{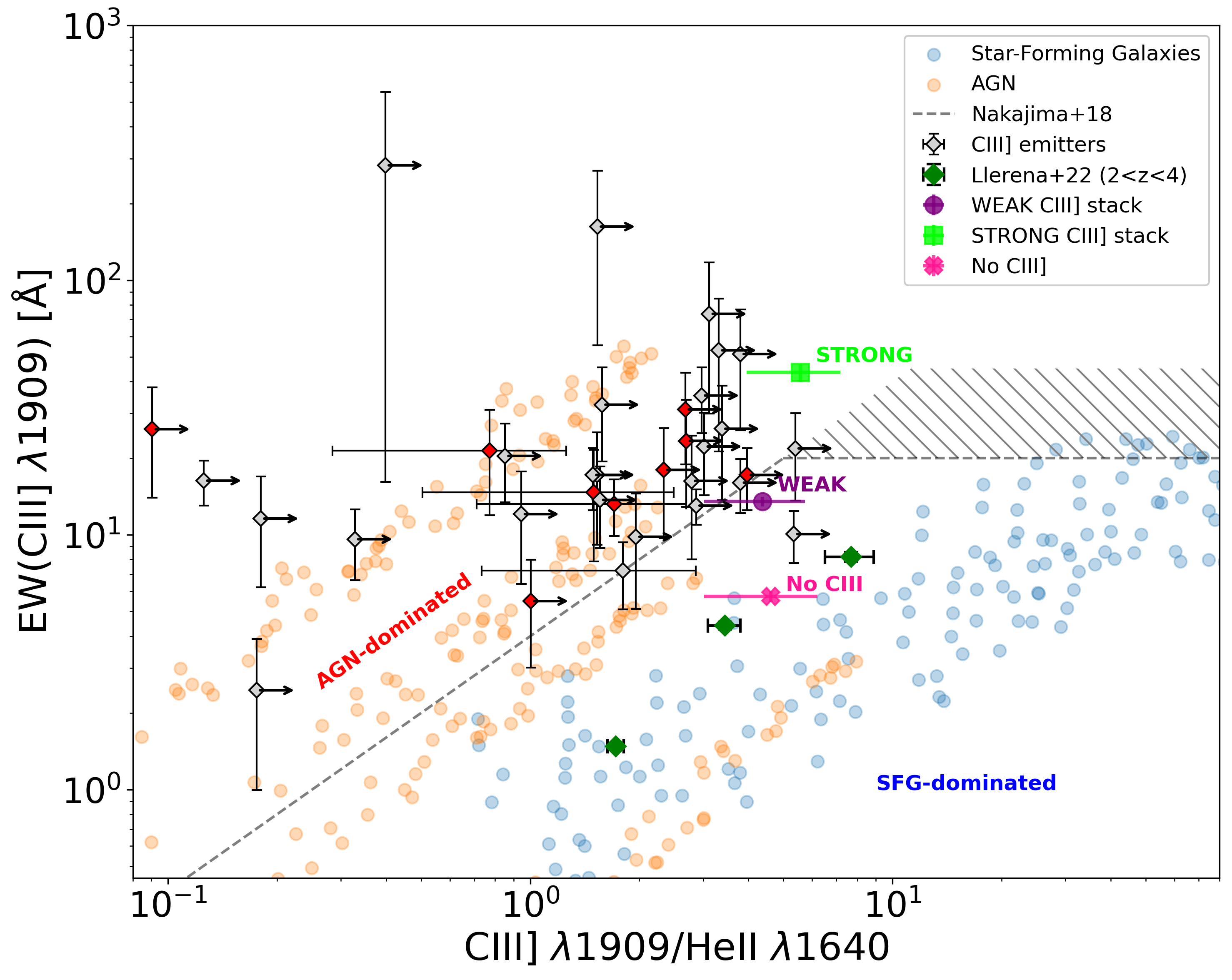}
    \caption{UV-diagnostic diagram using the EW of \ion{C}{III}] versus the ratio  \mbox{\ion{C}{III}]/\ion{He}{II} $\lambda$1640}. The SFG and AGN models are from \protect\cite{p11_NakajimaMaiolino2022} while the dashed black line is from \cite{p12_nakajima2018}. The values measured from the stacked spectra are shown as purple (Weak stack) light green (Strong stack) and pink (no \ion{C}{III}] stack) points, with lower-redshift stacks (darker green) from \cite{Llerena2022}. The \ion{C}{III}] emitters with a broad $\mathrm{H}\alpha$ are plotted as red diamonds.}
    \label{fig:CIII_UV_diagnostics}
\end{figure}

In Figure \ref{fig:CIII_UV_diagnostics}, we show the EW(\ion{C}{III}]) as a function of the \mbox{\ion{C}{III}]/\ion{He}{II} $\lambda$1640} ratio. From our sample of 39 galaxies, 4 exhibit a S/N$>2$ for the \ion{He}{II} $\lambda$1640 line, while the rest present upper limits of the \ion{He}{II} $\lambda$1640 line. Results for the stacks are also included in the plot.
Based on the commonly employed demarcation lines from \cite{p12_nakajima2018}, we find that 1 galaxy resides in the star-forming region. Another 22 galaxies occupy the AGN region: even when accounting for potential rightward shifts due to their upper limits on the x-axis, their high EW(\ion{C}{III}]) values would maintain their classification as AGN. Note that 4 of those galaxies fall outside the plotted range of the diagram. They present upper limits of the \ion{He}{II} $\lambda$1640 line and they are positioned in the left part of the graph. For clarity, and to improve visibility, we set axis limits that exclude this point from the displayed range.
Additionally, we identify 16 galaxies currently located in the AGN region that could potentially shift into the SFG region. For the purposes of our analysis, we classify them as potential AGN.
We also highlight in red the \ion{C}{III}] emitters that show a broad $\mathrm{H}\alpha$ component. 10 sources out of 14 could be plotted, where 7 of them are AGN and 3 of them are potential AGN according to the UV diagram. This shows a good agreement between the two diagnostics. 

An examination of the stacked spectra reveals the following  classifications: the strong \ion{C}{III}] stack clearly resides in the AGN region, while the weak \ion{C}{III}] stack lies within the SFG region but near the demarcation line. The stack of non-\ion{C}{III}] emitters is clearly located in the SFG region, very close to the stacked spectra of galaxies at lower redshift presented in \cite{Llerena2022}.

\begin{figure}[t!]
    \centering
    \includegraphics[width=\columnwidth]{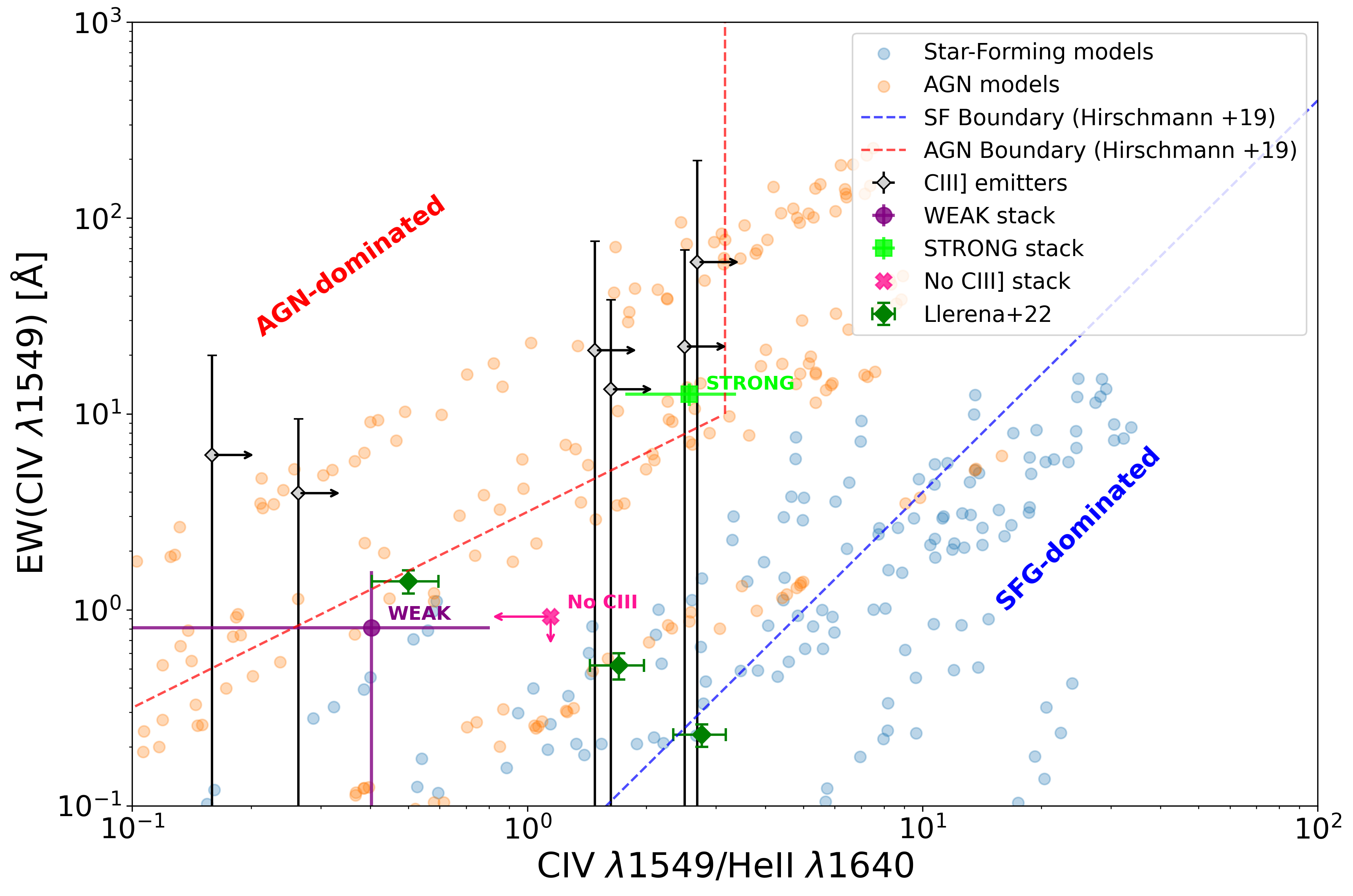}
    \caption{UV-diagnostic diagram using the EW of \ion{C}{IV} versus the ratio  \mbox{\ion{C}{IV}/\ion{He}{II} $\lambda$1640}. Demarcation lines given by \cite{2019MNRAS.487..333H}. Models and stacks same as in Figure \ref{fig:CIII_UV_diagnostics}.}
    \label{fig:CIV_UV_diagnostics}
\end{figure}

An alternative UV diagnostic plot involving EW(\ion{C}{IV}) vs \ion{C}{IV}/\ion{He}{II} ratio is shown in Figure \ref{fig:CIV_UV_diagnostics}. \ion{He}{II} $\lambda$1640 remains undetected in the 6 galaxies showing \ion{C}{IV} emission, thus yielding lower limits on the \ion{C}{IV}/\ion{He}{II} ratios. Although these 6 galaxies are all located in the AGN region of the diagram, they have lower limits in the x-axis (so they could shift to the mixed of SF dominated region) and they also exhibit large uncertainties  in the y axis. Overall we consider this analysis not conclusive for individual sources.
We also include the results from the stacks  as in previous analyses. The strong stack falls again well within the AGN region, while the weak stack occupies a mixed region containing both AGN and SFG models. The non-\ion{C}{III}] stack is also located in this mixed region, though the \ion{C}{IV} line remains undetected in this spectrum. 

\subsection{OHNO diagram}
\label{subsec:OHNO}

The OHNO diagram \citep{Backhaus_2022} has been proposed as a potential AGN diagnostic based on the ratios [\ion{Ne}{III}]/[\ion{O}{II}] and [\ion{O}{III}]/H$\beta$, which rely on four rest-optical emission lines that are relatively accessible at high redshift. A key advantage of the OHNO diagram over the traditional BPT diagram \citep{BaldwinPhillipsTerlevich_1981} is that it does not rely on H$\alpha$ and [\ion{N}{II}] $\lambda$6584, which can be blended together in low resolution spectra and fall outside the observable range of JWST at $z>7$.

While useful, the OHNO diagram is known to be more sensitive to the gas-phase metallicity and ionization parameter than to the fundamental source of the ionizing spectrum itself, as the chosen line ratios are largely degenerate in their ionization potentials \citep{2025ApJ_Cleri}. Therefore, while we use it for comparison, we place greater diagnostic weight on UV line ratios in our analysis. To explore the systematic uncertainties in this diagnostic, we examine how our results shift under different AGN and SFG models.

\begin{figure}[t!]
    \centering
    \includegraphics[width=\columnwidth]{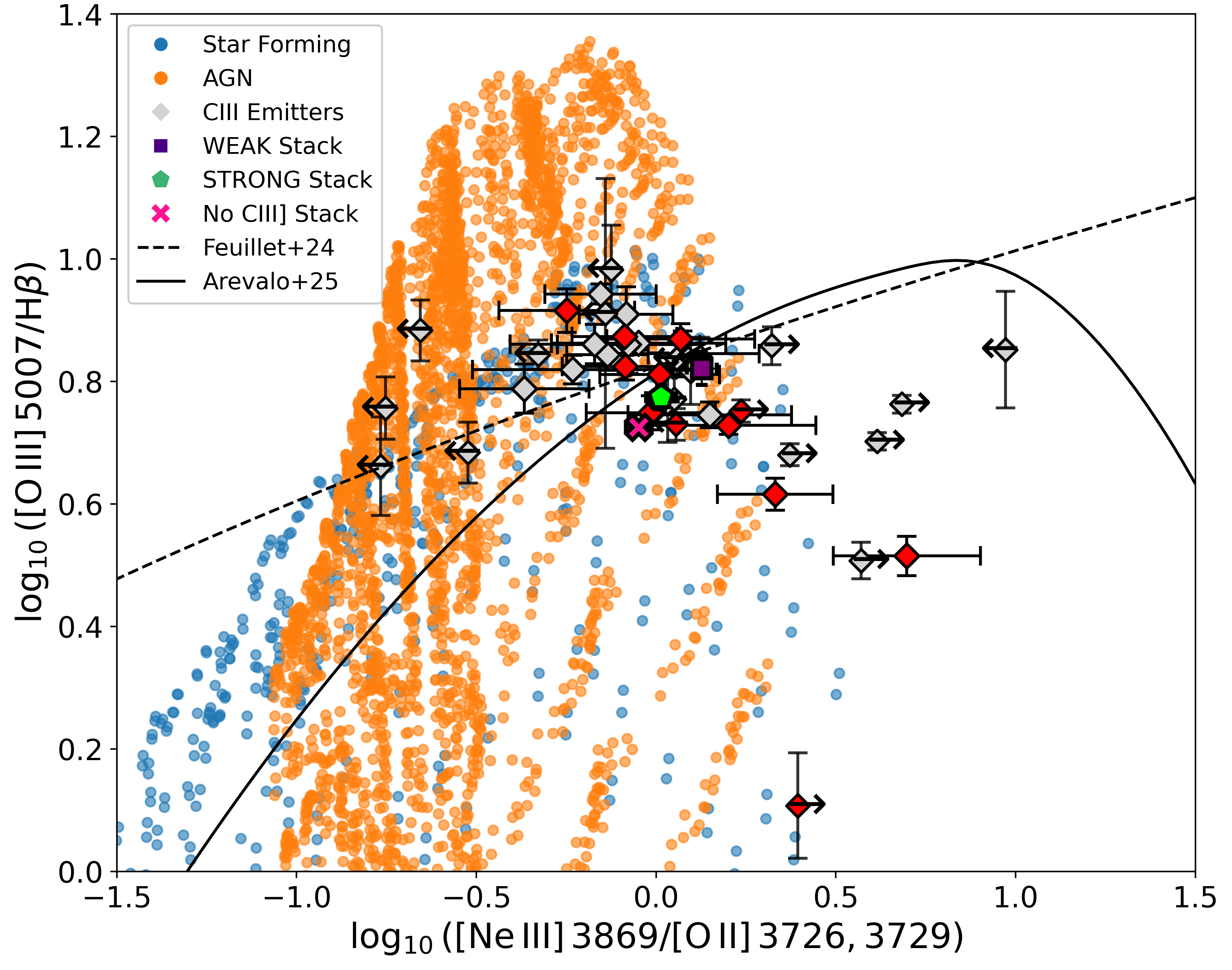}
    \caption{OHNO diagram displaying SFG models from \protect\cite{Gutkin2016} and AGN models from Plat al.(in prep.). The dashed black demarcation line is based on \protect\cite{Feuillet_2024} while the solid black line is based on \protect\cite{2025mnras_Arevalo}. The stacked measurements are shown as purple (Weak stack), light green (Strong stack) and pink (no \ion{C}{III}] stack) points. The \ion{C}{III}] emitters with a broad $\mathrm{H}\alpha$ are plotted as red diamonds.}
    \label{fig:OHNO_diagram}
\end{figure}

For 41 sources in our sample, we can measure or place meaningful limits on both line ratios and we present the results in Figure \ref{fig:OHNO_diagram}. Note that 1 galaxy falls outside the plotted range of the diagram. It is an upper limit positioned in the right part of the graph. For clarity, and to improve visibility, we set axis limits that exclude this point from the displayed range.

First, we plot the separation lines from \cite{Feuillet_2024}, derived empirically from SDSS spectroscopic data, and from \cite{2025mnras_Arevalo}  which is based on the previously mentioned photoionisation models of \cite{p11_NakajimaMaiolino2022}. We further plot SFG models from \cite{Gutkin2016}, and for AGN a new set of models from Plat et al. (in prep.), which are based on the recent AGN spectra of \cite{2018MNRAS_Kubota} and incorporate dependencies on black-hole mass and Eddington ratio. For both sets of models, we use an ionization parameter range of $-4 < \log U < -2$. The metallicity ranges are $0.0001 < Z < 0.03$ for the Plat et al. models and $0.0002 < Z < 0.03$ for the Gutkin et al. models. We also assume ranges of $3 < \log(M_{\rm BH}/M_\odot) < 9$ for black hole mass and $-1.5 < \log(\lambda_{\rm Edd}) < 2$ for the Eddington ratio. The Gutkin et al. models further assume an initial mass function (IMF) with an upper mass cutoff of 300 $M_\odot$ and a constant SFR with stellar population ages of 3 Myr and 10 Myr.

As shown in Figure \ref{fig:OHNO_diagram}, we identify 20 AGN (based on the \cite{Feuillet_2024} line) or 22 AGN (based on the \cite{2025mnras_Arevalo} line): according to those diagnostics, sources above the line can be classified as AGN while souces below the line could be either SFG or AGN origin.

When comparing our sources with the models by Plat and Gutkin, the situation is less clear, as  only 2 sources are in the AGN-only region while the rest of them being in the mixed region where we have both AGN and SFG models. This large discrepancy again showcases that the OHNO diagram is not as efficient in identifying AGN as the UV diagrams since the results heavily depend on the models or demarcation lines used. 

We also highlight in red the \ion{C}{III}] emitters that show a broad $\mathrm{H}\alpha$ component. Using the demarcation lines, of the 12 sources that can be placed on the diagram, 4 of these sources fall in the AGN region while the remaining 8 lie in the mixed region. All 12 lie in the mixed region when compared to the photoionization models by Plat+Gutkin. 

Upon examination of the stacked spectra, the three stacks fall below both diagnostic lines, placing them very close to each other in the mixed region. This again shows the low efficiency of the OHNO diagram compared to the UV diagrams that separate the stacks in the different regions. 

\subsection{ AGN component from SED fitting } \label{sec:AGN_comp}
An additional diagnostics to characterize our sources comes from 
SED fitting: in particular, we compare the quality of the best-fit SED photometric models with and without an AGN component, using their reduced chi-squared values. Photometric data were unavailable for six sources, so this comparison could only be performed for 55 sources in total. 
%The galaxy RUBIES-127820 could not be placed in any of the diagnostic plots and is therefore excluded from this analysis.

The results show that for 13 sources, the fit with the AGN component had a lower reduced chi-squared value, indicating that this component is needed to reproduce the photometry. Amongst those 13 sources, 11 present  other secure AGN indication either from the presence of a broad emission line component, or from the UV diagram or both. This set includes the identified 5 LRD (see next section). The remaining two sources for which the AGN component is needed, namely GDS-210003 and CAPERS-41455, are also potential AGN from the UV diagram.  
%For four of these LRDs, the difference in reduced chi-squared exceeds 10, indicating a significantly better fit when AGN component is included . 
%In particular, RUBIES-49140 exhibits an extreme difference of 2956.36 between the two models. 
%Given the complex spectral nature of the LRDs, we also performed a spectro-photometric fit for them. \textbf{do we have anything to say significant here ?do the values of masses and SF change with spectrophotometry}
For this set, we calculate the fractional contribution of the AGN component to the optical rest-frame continuum for 8 out the 13 cases (we did not include the LRDs results since a reliable AGN model has not been established yet). The AGN fraction is computed from the photometric spectrum at a rest-frame wavelength of 5500 \AA, chosen to avoid contamination from emission lines. We find an average AGN fraction of 62.3\% with a standard deviation of 26.3\%. 

For the other 42 sources, the SED fitting without the AGN component is preferred. Within this group, 25 have previous AGN indications (15 confirmed AGNs and 10 potential AGNs). However, for most of them (14 confirmed and 7 potential AGNs), the difference in the reduced $\chi^2$ between the two fits is actually rather small, indicating that the SED fitting does not strongly prefer either model. On the other hand, for the only galaxy classified as a secure SFG in the UV diagram, namely GDS-202208, we find  a very  large reduced chi-squared difference of 160, strongly favoring the model without an  AGN component, in agreement with the emission line diagnostics. Finally, we note that the galaxy RUBIES-127820, which could not be placed on previous diagnostic plots, is also best fit by a model without an AGN, though with a small chi-squared difference.

We conclude that SED fitting based on photometry alone securely identifies approximately one third of our AGN population. This fraction could be significantly improved by incorporating spectro-photometric SED fitting. 
Based on our analysis, when a source is classified as an AGN by SED fitting, the identification is reliable, and no star-forming galaxies are misclassified as AGN.

\section{Are there LRDs amongst \ion{C}{III}] emitters?}\label{sec:LRDs}

In just less than 3 years of operations, the JWST has provided us with many surprises. Amongst them, the discovery of a completely new population of red, compact sources that are unexpectedly abundant in the early phase of our Universe, the so-called Little Red Dots (LRDs) \citep[e.g.,][]{matthee2024littlereddotsabundant,2024ApJ...968....4P}.
LRDs are very compact sources, with a characteristic V-shaped SED and, in the majority but not all cases, with broad hydrogen emission lines with FWHM$\approx 2$-3 $10^3 ~\rm km ~s^{-1}$ that have been interpreted as evidence of AGN activity \citep[see e.g.][]{kocevski2025risefaintredagn}. 
%In the rest frame the apex of the V is typically observed around 3.55 \AA, the "blue arm" of the SED has a blue color, ie. increasing toward shorter wavelengths, while the "red arm" is correspondingly red and exhibiting a pronounced "Balmer Break".  
%red-optical and blue-UV sources representing a potentially crucial phase of early black hole growth.  Characterized by “V-shaped” spectral energy distributions and broad emission lines, they appear to be faint AGN, often heavily obscured. 
Indeed, current analyses \citep[e.g.,][]{kocevski2025risefaintredagn} reveal that over 80\% exhibit broad-line features when brown dwarf contaminants are excluded. One of the latest models that seems to reproduce most of the observed spectro-photometric features involves the idea of a BH$^*$ as the engine of LRDs, i.e., a BH surrounded by a thick dense gas envelope \citep{inayoshimaiolino2025, Naidu2025A, degraaff2025b}. Despite these advances, fundamental questions persist regarding the physical nature of their obscuring material and their overall contribution to cosmic evolution. 

Following the classification criteria from \cite{kocevski2025risefaintredagn} and \cite{barro2024}, which use photometric data to classify sources as LRD candidates, we check the presence of LRDs in our sample of \ion{C}{III}] emitters. A total of 9 sources pass this photometric selection. To secure the LRD sample, we also use the available spectroscopic data of these sources to estimate both the ultraviolet and optical slope, masking emission lines. We use a Markov Chain Monte Carlo (MCMC) approach to fit the slopes modelled as power laws. 

%For those galaxies that pass the slope requirements (having a UV slope $-2.8 < \beta_{\mathrm{UV}}< -0.37$ and an optical slope $\beta_{\mathrm{opt}}>0$), we check the compactness of the source.

\begin{figure*}[t]
    \centering
    
    % Top row with two figures
    \includegraphics[width=0.40\textwidth]{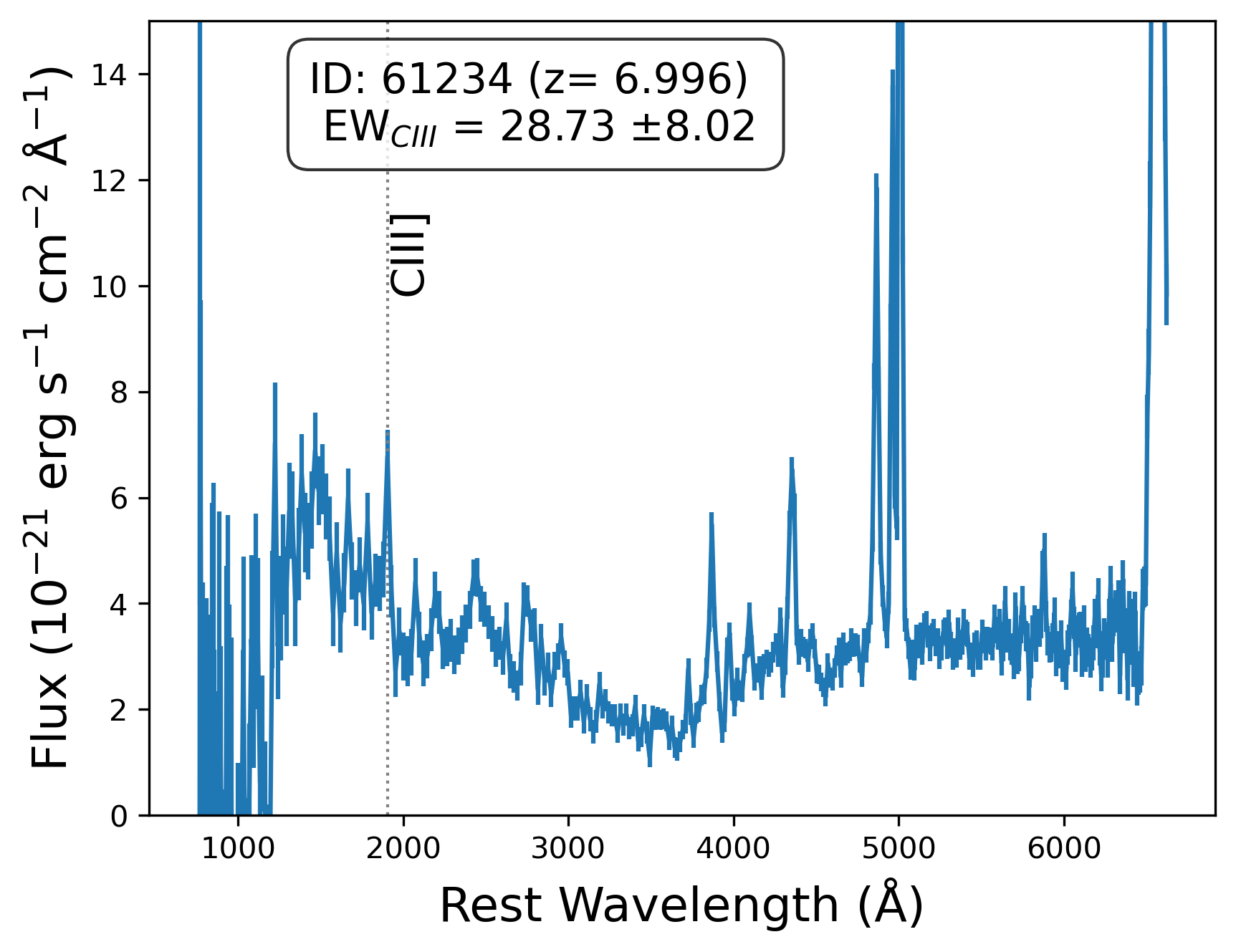}
    \includegraphics[width=0.40\textwidth]{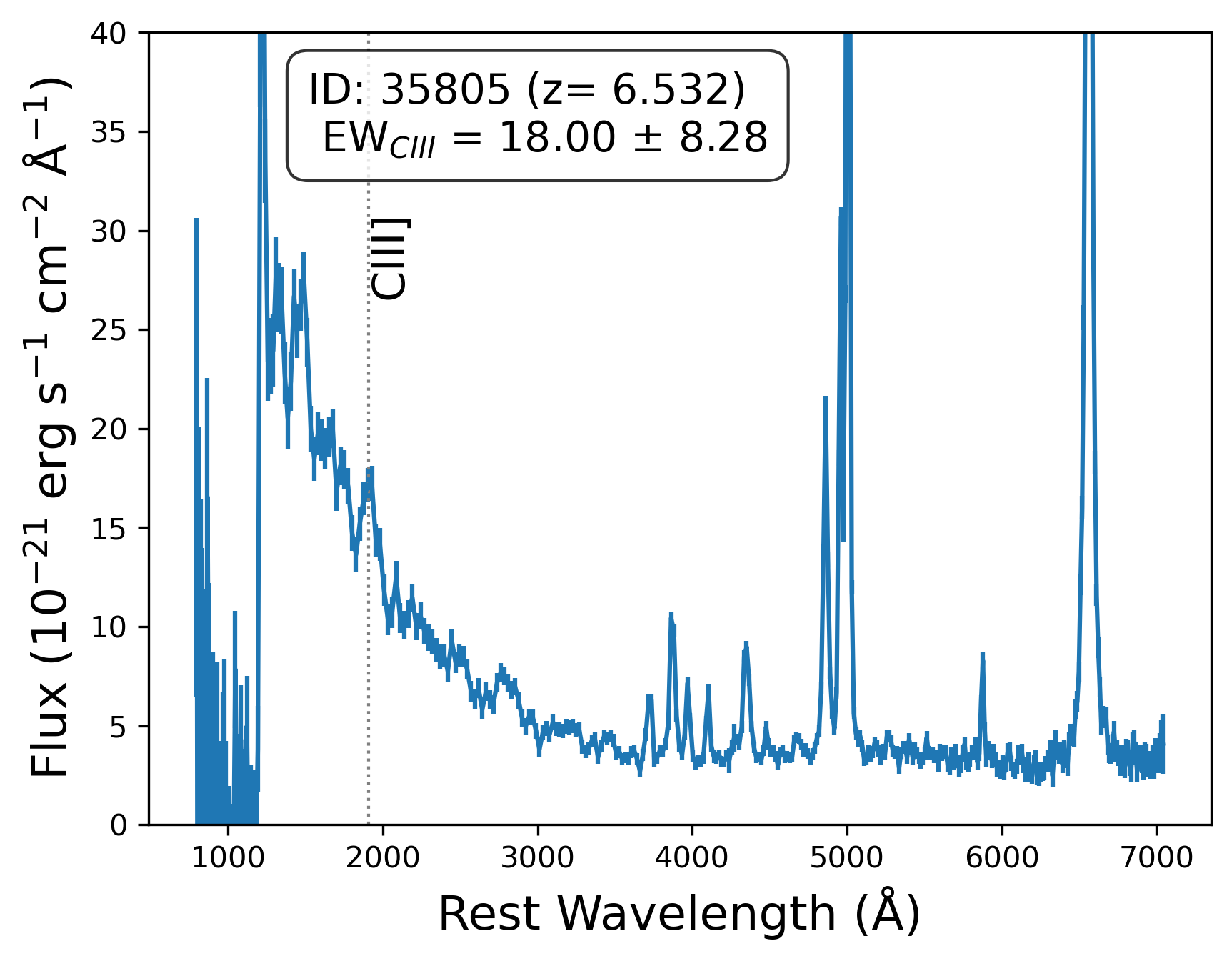} \\
    
    % Bottom row with three figures
    \includegraphics[width=0.33\textwidth]{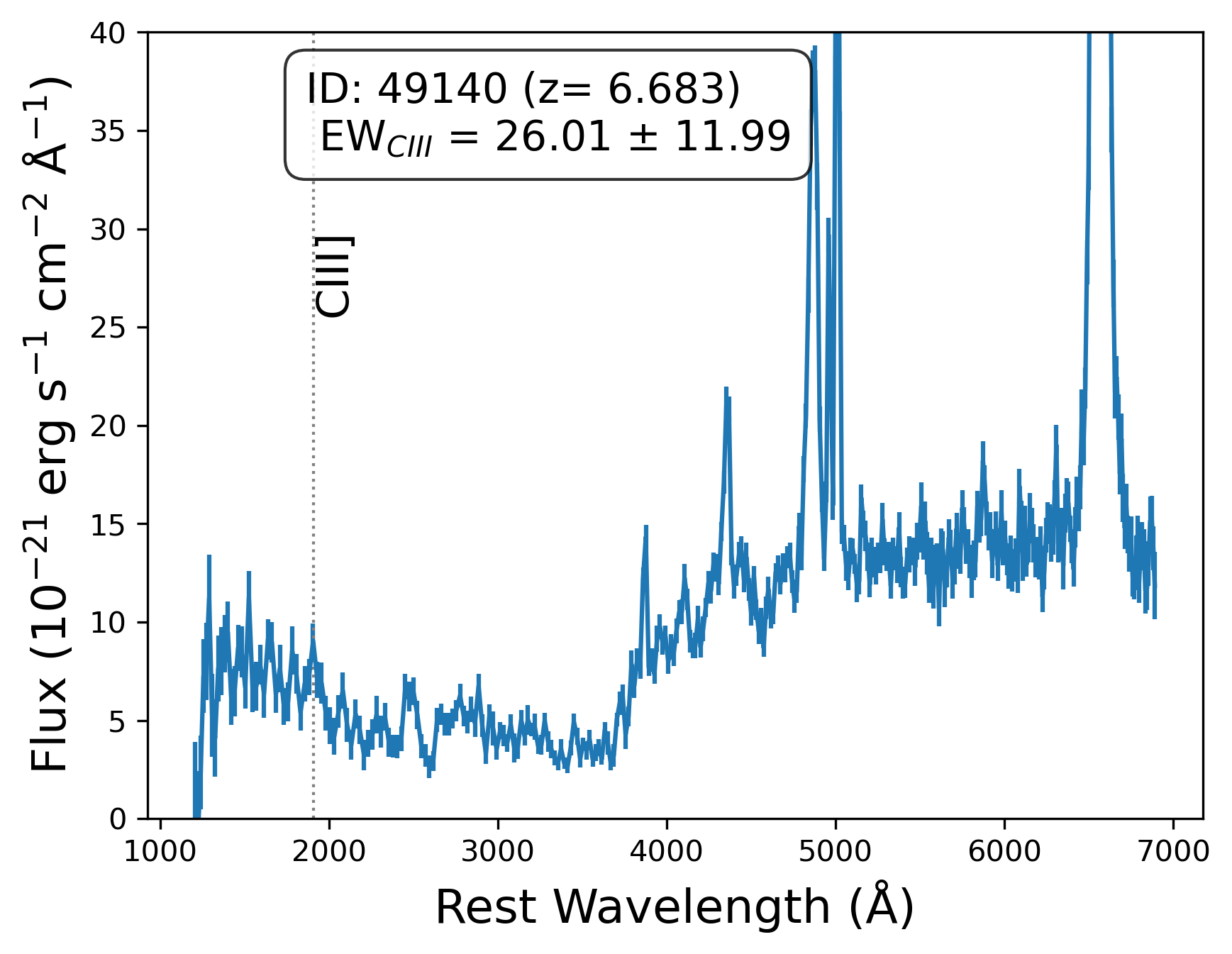}
    \includegraphics[width=0.33\textwidth]{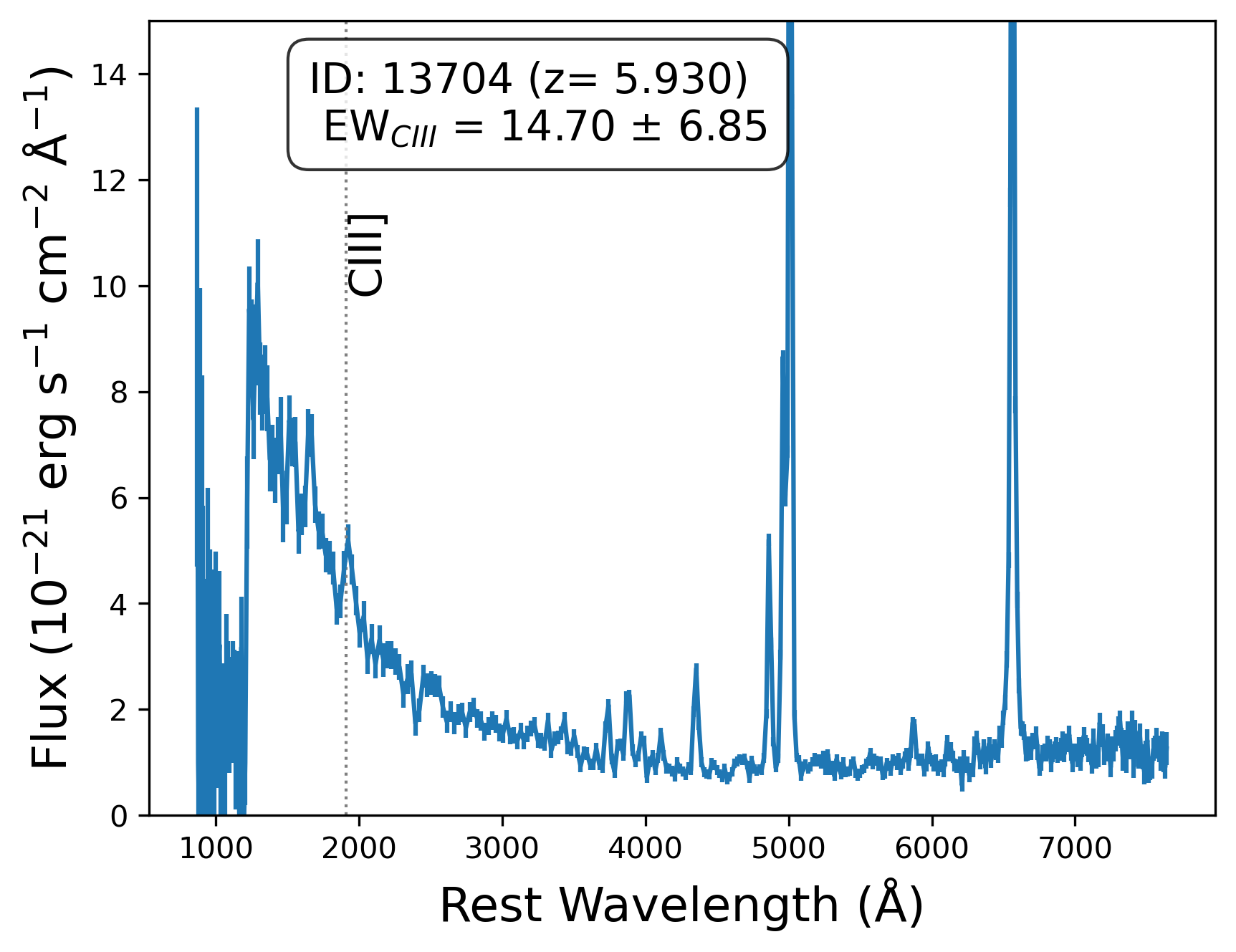}
    \includegraphics[width=0.33\textwidth]{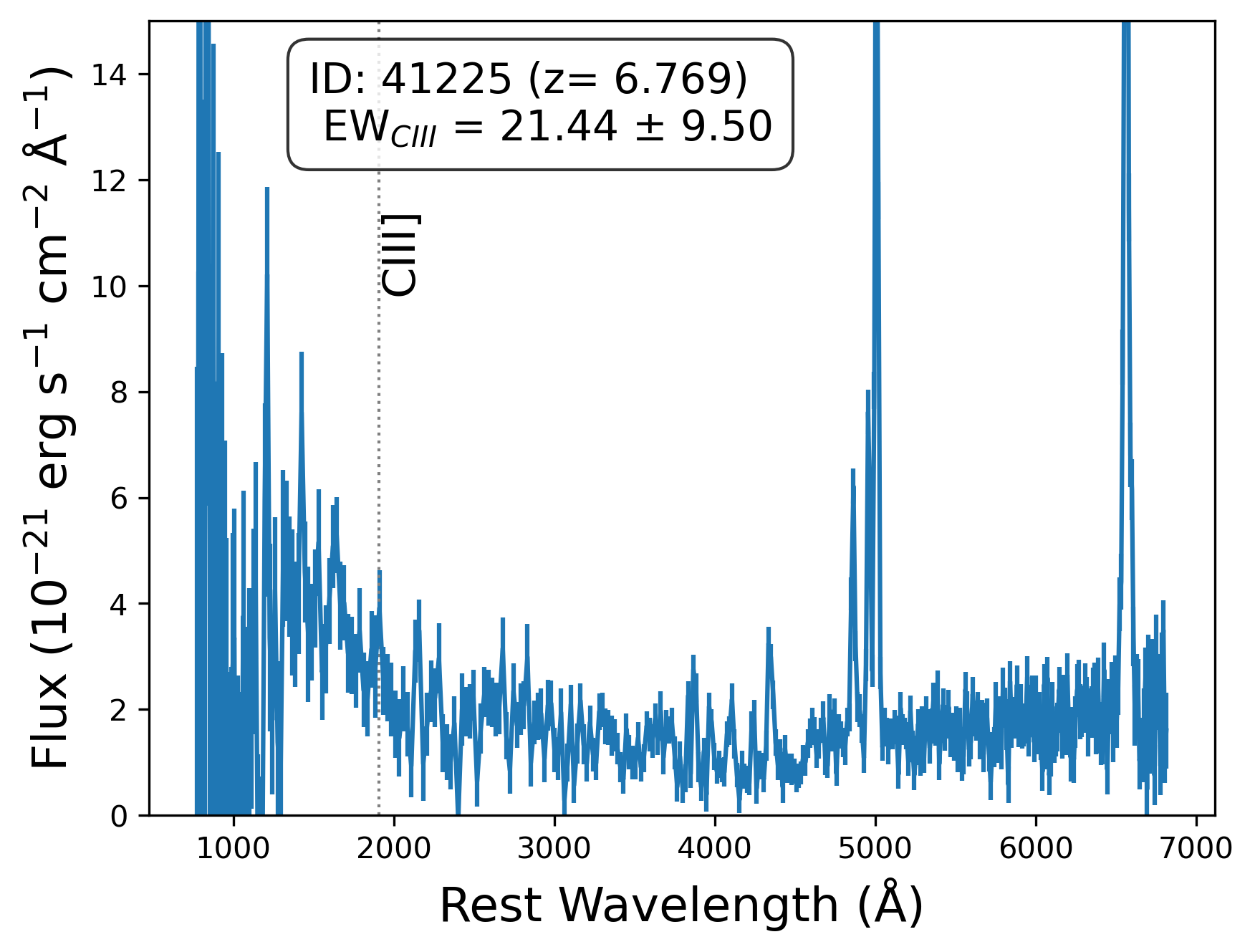}
    
    \caption{Rest-frame spectra of the five Little Red Dots (LRDs) identified in our sample. From left to right, top to bottom: COSMOS-61234, CAPERS-35805, RUBIES-49140, GDS-13704, and UNCOVER-41225. }
    \label{fig:LRD_figure}
\end{figure*}

We confirme 5 of 9 LRD candidates that passed the color requirements spectroscopically, having a spectroscopic UV slope $-2.8 < \beta_{\mathrm{UV}}< -0.37$ and an optical slope $\beta_{\mathrm{opt}}>0$. Four of them were already identified in previous papers as LRDs, and is a new identification  from the CAPERS survey. The spectra of these candidates are shown in Figure \ref{fig:LRD_figure}. We find that standard SED fitting with Bagpipes using photometric data alone fails to accurately reproduce the observed photometry of these LRDs. Although spectrophotometric fitting provides improved fits, purely stellar models remain inadequate. Although there is considerable effort to understand the characteristics of LRDs \citep[e.g.,][]{2025arXiv_Inayoshi}, currently SED fitting tools (e.g. Bagpipes, Prospector, Beagle, Galfit) lack appropriate AGN models tailored for LRDs, preventing reliable mass estimates. Because of this, we exclude these sources from the Main Sequence diagram, as their derived stellar masses are still unreliable.

%Comparing with other spectroscopically confirmed LRDs, 
We note that among the LRDs known at $z > 7$ in the literature, two exhibit \ion{C}{III}] emission \citep[e.g.,][]{2025arXiv_Jones, 2025arXiv_Morishita}. Additionally, a LRD at $z=5.3$ has been reported to show  \ion{C}{III}] based on new spectroscopic observations of the Bullet Cluster \citep[e.g.,][]{tripodi2025}, which were not available at the beginning of our analysis.

Overall the LRD fraction in our sample of \ion{C}{III}] emitters is low, less than 10\%: this together with the relatively sparse presence of \ion{C}{III}] in LRD spectroscopic samples (around 20\% and mostly in redder LRDs,  Barro et al., in preparation), confirms that while the \ion{C}{III}] emitters and LRD are both AGN dominated, they trace  distinct populations.

\section{The nature of \ion{C}{III}] emitters}\label{sec:Nature_CIII_emitters}

To assess the presence of AGN within our sample of \ion{C}{III}] emitters, we employ the combination of the complementary diagnostics described in the previous sections.
Broad H$\alpha$ emission is identified in 14 sources, which is of course a lower limit given that we are using prism spectroscopic observations. These are secure AGNs. From the UV diagnostics involving \ion{C}{III}], we find 22 galaxies securely classified as AGNs, out of a total of 39 galaxies which can be meaningfully placed  in the diagram, with another 16 identified as potential AGNs: all sources with broad H$\alpha$ that could be placed in the \ion{C}{III}] diagnostics have a consistent AGN classification. The \ion{C}{IV}-based diagnostic is unfortunately not so useful due to the limited number of detections, which are however all placed in the AGN dominated region. Noteworthy, The \ion{C}{IV}-based diagnostic is in agreement with the \ion{C}{III}] based one.

%Based on the OHNO diagram, we identify 22 AGN (using the demarcation line of protect\cite{2025mnras_Arevalo}) out of the 41 sources for which both line ratios could be measured. 

%, although the NIRSpec/PRISM configuration limits our ability to detect broad components in all sources.

Combining all diagnostics, we find that half of the sources in our sample---29 out of 61 galaxies—--exhibit at least one secure indication of AGN activity, with an additional 13 being potential  AGNs, based on the \ion{C}{III}] diagnostic. 
These factions are much higher that the typical AGN fraction in the general population at similar redsihft \citep{juodžbalis2025jadescomprehensivecensusbroadline}.

A summary of these diagnostics is provided in Table~\ref{tab:CIII_table}. This group also includes the five sources identified as LRDs, all of which exhibit broad-line components. Although we have demonstrated that the OHNO diagram is highly unreliable, we note that most of the sources that would be identified as AGN according to the \cite{2025mnras_Arevalo} demarcation line, are also AGN according to the other  more secure diagnostics. 

Examining the stacked spectra, we find that in the OHNO diagram, all three stacks fall slightly below both demarcation lines, placing them in the mixed region and very close to each other. As expected, the UV diagnostics display more distinct behavior. As shown in Figure~\ref{fig:CIII_UV_diagnostics}, the strong \ion{C}{III}] stack clearly lies within the AGN region, while the weak \ion{C}{III}] stack occupies the star-forming region but remains close to the demarcation line. The stack of non-\ion{C}{III}] emitters is located well within the star-forming region, in good agreement with the stacked spectra of lower-redshift galaxies from \citet{Llerena2022}. Similarly, in Figure~\ref{fig:CIV_UV_diagnostics}, the strong stack is again within the AGN region, the weak stack occupies a mixed zone containing both AGN and SFG models, and the non-\ion{C}{III}] emitters remain consistent with SFGs, though plotted as upper limits given the non-detection of the \ion{C}{IV} line.

We also show in Sect. \ref{sec:AGN_comp} that SED fitting using photometry alone is able to identify robustly about a third of the AGN that are identified spectroscopically.  This results could be probably improved by performing spectro-photometric analysis.

Overall, our \ion{C}{III}]–emitting sources are predominantly AGN. This naturally accounts for the EWs observed compared to those measured in cosmic noon galaxies with similar $M_{\rm UV}$ in the VANDELS survey \citep{Cunningham_2024,Llerena2022}, which is known to be dominated by star-forming systems. A VANDELS-like population of galaxies with modest \ion{C}{III}] emission is likely also present at high redshift; indeed, the stacked spectrum of galaxies without individual line detections shows weak \ion{C}{III}] emission (with EW(\ion{C}{III}]) $\simeq 5 \AA$) and lies consistently within the star-forming regions of all diagnostic diagrams explored.

Our JWST high-redshift sample also includes a subset of star-forming galaxies exhibiting high EWs(\ion{C}{III}]) that are not represented in the VANDELS sample. This discrepancy may reflect differences in ISM conditions at very high redshift, such as harder ionizing radiation fields, lower metallicities, or higher electron densities \citep[e.g.,][]{Llerena2022,p12_nakajima2018}.

\section{Summary and conclusions} \label{sec:Conclusion}
We present a systematic study of \ion{C}{III}] emission in galaxies at redshifts $5 < z < 7$, using JWST/NIRSpec prism spectroscopy. From a  parent sample of 1896 galaxies, we identify 61 robust \ion{C}{III}] emitters. Our analysis involved measuring emission line fluxes and equivalent widths using \lime and deriving physical properties (e.g., stellar masses, star formation rates) through SED fitting with \textsc{Bagpipes}. Our approach reveals a sample of  high-z \ion{C}{III}] emitter population that  exhibit significantly higher rest-frame equivalent widths compared to \ion{C}{III}] emitters at lower redshifts with a similar $M_{\rm UV}$ range (see Figure \ref{fig:EW_CIII_vs_Muv}): while  the prism observations we are using do not allow us to probe emission lines with low EW due to the low resolution, we also note that none of the intermediate redshift VANDELS sources exhibit EW higher than $\simeq 10-15 \AA$. This results in the small  overlap between the two samples.

Our \ion{C}{III}] emitters span the full range of the star-forming main sequence, with no substantial difference in position from the parent population.
%which could suggest they represent a diverse population of galaxies hosting a central AGN in this epoch. Those below the main sequence with high EWs may represent galaxies where AGN feedback is beginning to suppress star formation. 
To understand the nature of our \ion{C}{III}] emitters, and in particular the difference with the lower redshift sample, we employed a number of different AGN diagnostics, including optical and UV emission lines.
Such diagnostics indicate that AGN activity is common among our \ion{C}{III}] emitters. 
Broad emission lines are detected in 14 objects, but the limited spectral resolution of NIRSpec/prism reduces our sensitivity to such features, implying that this represents a lower limit on the true number of broad-line sources. The \ion{C}{III}] diagnostics also identifies 22 sources as secure AGN and 13 additional one as potential AGN; the CIV diagnostics although limited to few sources, also identifies 6 potential AGN. We have also employed the ONHO diagram, which involves optical lines, however it resulted to be highly unreliable as models gave inconsistent classifications. 
Finally, SED fitting using photometry alone  is able to identify robustly about a third of the AGN that are independently identified spectroscopically.
Altogether, at least half of our sample show  one or more AGN signature, and including the potential AGN.
Stacked spectra reinforce these trends: the strong \ion{C}{III}] stack occupies the AGN region for the UV diagnostics, while the weak and non-\ion{C}{III}] stacks align with SFG models.

Our results suggest that one possible reason for the difference in EW between the lower redshift sample from \citep{Cunningham_2024} and higher redshift samples from our work could be indeed the presence of AGN in the majority of our sample. 
%Furthermore, galaxies that are not AGN but still exhibit high equivalent widths may be attributed to harder radiation fields, lower metallicities, and/or higher ionization parameters. 
Higher-resolution and/or deeper follow-up observations are therefore crucial to robustly disentangle the two populations. Such data would enable the detection of broad components in additional Balmer lines and of fainter AGNs, improve the completeness limit of \ion{C}{III}], and resolve the \ion{He}{II}–\ion{O}{III}] $\lambda1666$ blend, thereby strengthening the reliability of UV diagnostic methods. Moreover, these observations would allow us to probe the physical nature of galaxies not currently classified as AGNs yet exhibiting high equivalent widths, clarifying whether their properties arise from harder ionizing spectra, lower metallicities, higher electron densities, and/or elevated ionization parameters—scenarios that can be directly tested with targeted follow-up.
\begin{acknowledgements}
%We thank the anonymous referee for the constructive feedback provided. 
This work is based on observations made with the NASA/ESA/CSA James Webb Space Telescope, obtained at the Space Telescope Science Institute, which is operated by the Association of Universities for Research in Astronomy, Incorporated, under NASA contract NAS5-03127. These observations are associated with programs \#6368, \#4233, \#1215, \#2565, and \#3543.
Support for program number GO-6368 was provided through a grant from the STScI under NASA contract NAS5-03127. The data were obtained from the Mikulski Archive for Space Telescopes (MAST) at the Space Telescope Science Institute. These observations can be accessed via \href{http://dx.doi.org/10.17909/0q3p-sp24}{DOI}. MLl acknowledges support from the INAF Large Grant 2022 “Extragalactic Surveys with JWST” (PI L. Pentericci), the PRIN 2022 MUR project 2022CB3PJ3 - First Light And Galaxy aSsembly (FLAGS) funded by the European Union – Next Generation EU, the INAF Mini-grant 2024 "Galaxies in the epoch of Reionization and their analogs at lower redshift" (PI M. Llerena), and the Large Grant RF 2023 F.O. 1.05.23.01.11 "The MOONS Extragalactic Survey". RA acknowledges support of Grant PID2023-147386NB-I00 funded by MICIU/AEI/10.13039/501100011033 and by ERDF/EU and from the Severo Ochoa award to the IAA-CSIC  CEX2021-001131-S. JSD acknowledges the support of the Royal Society through the award of a Royal Society Research Professorship.

%We thank --- for kindly providing tabulated data from ---
\end{acknowledgements}

\bibliographystyle{aa}
\bibliography{biblio}

\begin{appendix}

\section{Slit Loss Corrections}
\label{sec:corr_factor}

\begin{figure}[t!]
    \centering
    \includegraphics[width=\columnwidth]{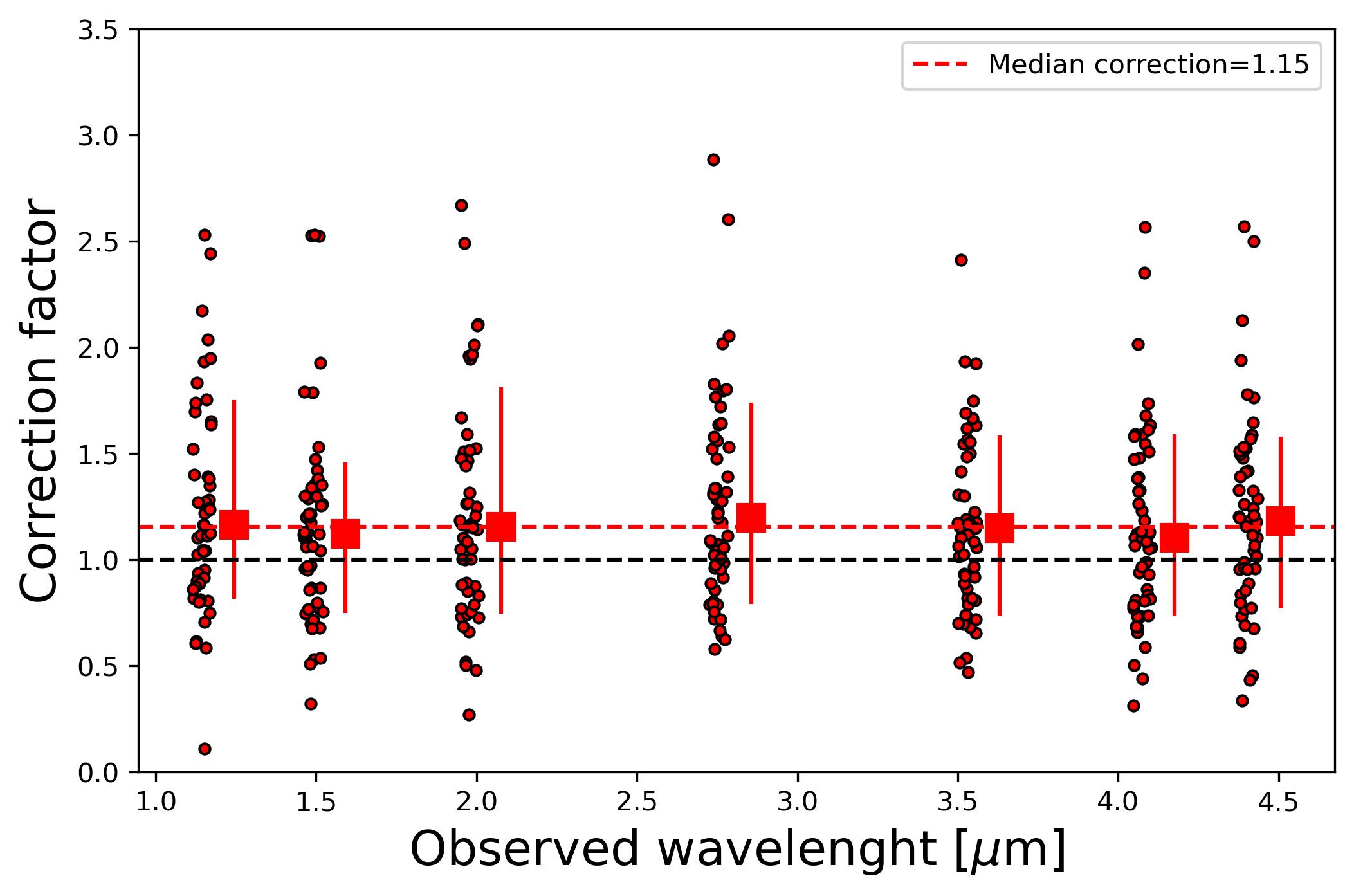}
    \caption{The correction factor versus observed wavelength for the sample of \ion{C}{III}] emitters. The median correction of 1.15 is shown by  the dashed red line}
    \label{fig:correction_factor}
\end{figure}

In Figure \ref{fig:correction_factor} we show the correction factor calculated for several filters as explained in \ref{sec:lime}. 

On average, there is a slit-loss factor of around 1.1-1.2, and it is wavelength-independent i.e.,  on average, we find the same correction factor across all filters although the scatter is large. For the galaxies where we find an average correction factor >1 (33 sources in total), the correction is made by multiplying the flux densities of the spectrum by the average factor for each galaxy. For galaxies with an average correction factor < 1, no correction is applied (meaning a correction factor = 1 is assumed) because in those cases there are no slit-losses.

% \section{Model Selection for Stellar Mass Measurement}
% \label{sec:model_selection}
% Here we describe how the best-fit model for the stellar mass is selected. For each galaxy, we compare the reduced chi-squared ($\chi^{2}{\mathrm{red}}$) values from fits performed with and without an AGN component. The preferred model is chosen by balancing the statistical improvement in $\chi^{2}{\mathrm{red}}$ with the other AGN diagnostics discussed in this paper.
% If the fit without the AGN component yields a lower $\chi^{2}_{\mathrm{red}}$ and no AGN signature is present, we adopt the stellar mass from that fit, i.e. based on galaxy models only. If the fit including the AGN component is statistically superior to the no-AGN one, and an AGN signature is confirmed in the source, we adopt the resulting stellar mass from the AGN fit. 
% In cases where the results are conflicting, i.e. when the no-AGN fit has a lower $\chi^{2}_{\mathrm{red}}$ but an AGN signature is present, we evaluate the difference in reduced chi-squared ($\Delta\chi^{2}_{\mathrm{red}}$). 

%A small difference ($\Delta\chi^{2}_{\mathrm{red}} < 10$) leads us to compare the stellar masses from both fits and if they are consistent, the mass from the no-AGN fit is adopted. 

%A large difference ($\Delta\chi^{2}_{\mathrm{red}} \geq 10$), however, results in a re-analysis of these galaxies using spectrophotometric data to better constrain the fit with an AGN component. 

\section{The \mbox{\ion{O}{III}] $\lambda$1665} / \mbox{\ion{He}{II} $\lambda$1640} Ratio}
\label{sec:ratio_calibration}

\begin{figure}[t!]
    \centering
    \includegraphics[width=\columnwidth]{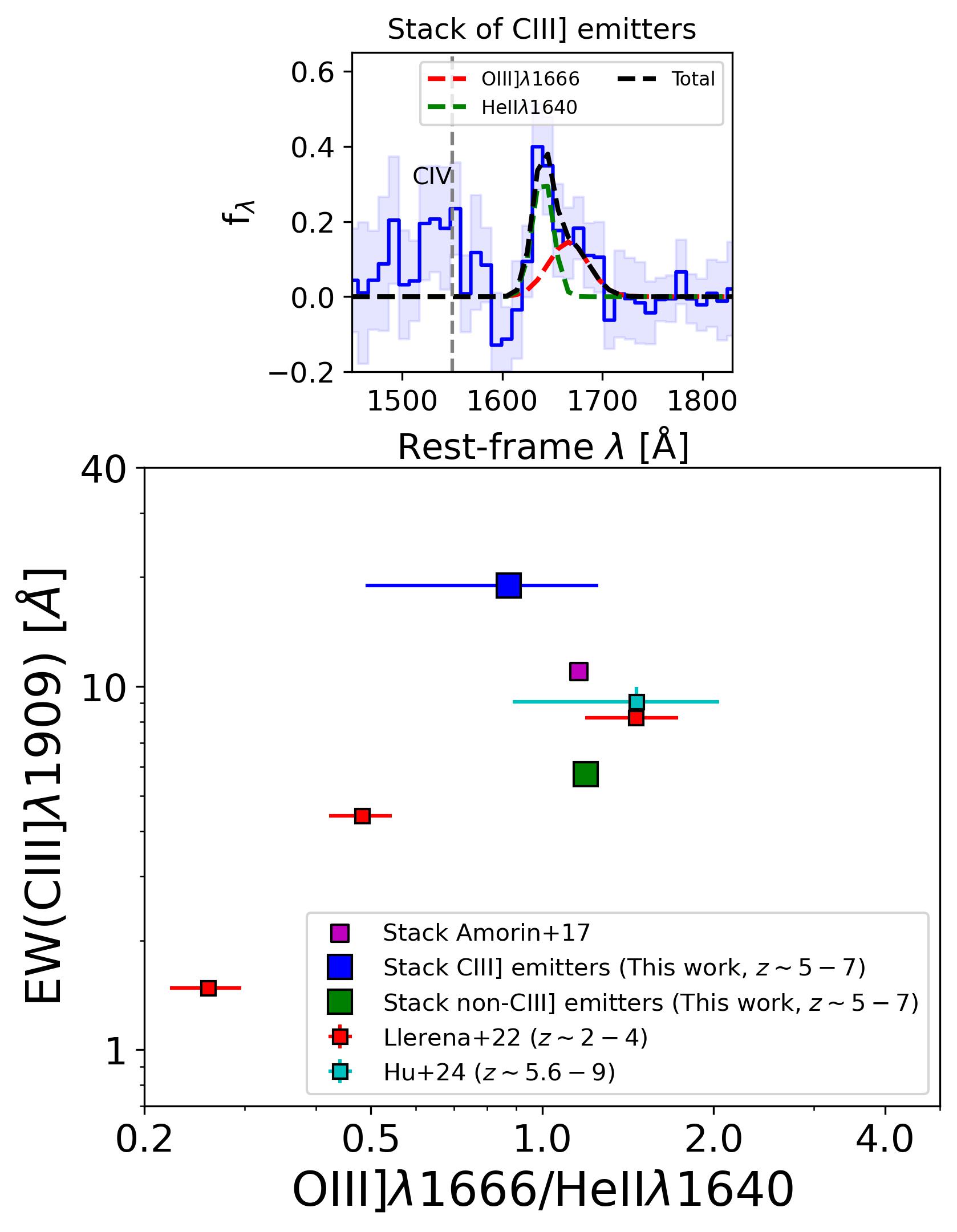}
    \caption{
    Top panel: Continuum-subtracted stacked spectrum of \ion{C}{III}] in the sample (blue line) and the corresponding uncertainty (blue shaded region). The dashed red and green lines represent the best-Gaussian fit of \mbox{\ion{O}{III}] $\lambda$1665} and \mbox{\ion{He}{II} $\lambda$1640}, respectively, while the dashed black line represents the total fit of the blend lines. The vertical gray line indicates the position of \ion{C}{IV}. Bottom panel:  The \mbox{\ion{O}{III}] $\lambda$1665} / \mbox{\ion{He}{II} $\lambda$1640} line ratio as a function of \ion{C}{III}] equivalent width. The blue square represents the \mbox{\ion{O}{III}] $\lambda$1665} / \mbox{\ion{He}{II} $\lambda$1640} ratio measured in the stack, yielding a value of 0.87$\pm$0.38. This value is compared to measurements from \citet{Amorin2017}, \citet{Llerena2022} and \citet{Hu2024}. }
    \label{fig:UV_diagnostic_stacks_o3}
\end{figure}

To obtain a reliable measure of the \mbox{\ion{He}{II} $\lambda$1640}  that we employ in the UV diagnostic diagrams, we 
calculate the \mbox{\ion{O}{III}] $\lambda$1665} / \mbox{\ion{He}{II} $\lambda$1640} line ratio from stacked spectra of our \ion{C}{III}] emitters. Following the methodology described in Sec. \ref{sec:stacking_method}, we stacked the spectra of the entire sample of selected \ion{C}{III}] emitters. Due to the low-resolution of the prism spectrum, the two lines are blended. However, the high SN of the stack allows up to model the two components separately and infer their relative contribution to the blend. For the modeling, we assumed two blended Gaussian components with fixed central wavelength for each line and a local continuum. In the top panel of Fig. \ref{fig:UV_diagnostic_stacks_o3} we show the results of the Gaussian modeling and the continuum-subtracted stacked spectra: clearly two Gaussians can fit the line blend much better than one component alone.  We measure a ratio of 0.87$\pm$0.38 in the entire stack of \ion{C}{III}] emitters and we thus apply this value to individual emitting galaxies. We repeated the exercise for stack of the non-\ion{C}{III}] emitters and the ratio derived is slightly higher but still well within the uncertainty. 
As shown in bottom panel in Fig. \ref{fig:UV_diagnostic_stacks_o3},  these empirically derived values  fall within the range observed for SFG in the VANDELS survey, showing particular consistency with the most extreme \ion{C}{III}] emitters (EW $\sim$5--10\,\AA) in that sample. Furthermore, the ratio ares consistent with the value derived from the stack presented in \citet{Amorin2017}, which itself occupies the AGN region of standard UV diagnostic diagrams. Finally very similar \mbox{\ion{O}{III}] $\lambda$1665} / \mbox{\ion{He}{II} $\lambda$1640} ratios are found when stacking galaxies at $z\sim5.6-9$ using NIRSpec medium resolution spectra \citep{Hu2024}. 

\section{Properties of the \ion{C}{III}] Emitters Sample}
\label{sec:table_data}

In Table \ref{tab:CIII_table}, we present the properties of the \ion{C}{III}] emitter sample, including coordinates, redshift, and the results of our AGN diagnostics. It also includes the EW values and the UV $\beta$ slopes. 

Table \ref{tab:stack_all_lines} shows the values of the line ratios and EWs used for our three stacks.

\twocolumn
\onecolumn
{\tiny
\begin{longtable}{lcccccccc} 

\caption{This table catalogs the properties of the \ion{C}{III}] emitter sample, including AGN diagnostics, emission-line equivalent widths and UV $\beta$ slopes. The galaxies below the dotted line are those for which we do not have the photometry. A -- means that the galaxy could not be placed on the plot.} \\
\label{tab:CIII_table} \\
\hline\hline
ID & RA [deg] & DEC [deg] & $z_{\mathrm{spec}}$ & Broad & UV diagnostic & $EW_{\mathrm{CIII]}}$ [\AA] & $\beta_{UV}$ \\
\hline
\endfirsthead

\multicolumn{9}{c}{{\tablename\ \thetable{} -- continued from previous page}} \\ % Also change this to 9!
\hline\hline
ID & RA [deg] & DEC [deg] & $z_{\mathrm{spec}}$ & Broad & UV diagnostic & $EW_{\mathrm{\ion{C}{III}]}}$ [\AA] & $\beta_{UV}$ \\
\hline
\endhead

\hline \multicolumn{8}{r}{{Continued on next page}} \\
\endfoot

\hline
\endlastfoot

CAPERS-121916 & 34.23591 & -5.08822 & 6.011 & Not broad & AGN & 73.73 $\pm$ 43.75 & $-1.36^{\genfrac{}{}{0pt}{}{-0.98}{+0.96}}$ \\
CEERS-397 & 214.8362 & 52.88269 & 6.007 & AGN & Potential AGN & 5.50 $\pm$ 2.48 & $-1.75^{\genfrac{}{}{0pt}{}{-0.08}{+0.07}}$ \\
COSMOS-61234 & 150.1069 & 2.360046 & 6.9959 & AGN & -- & 28.73 $\pm$ 8.02 & $-1.16^{\genfrac{}{}{0pt}{}{-0.22}{+0.22}}$ \\
COSMOS-67506 & 150.1266 & 2.39305 & 5.8666 & Not broad & AGN &  21.85 $\pm$ 8.25 &  $-1.59^{\genfrac{}{}{0pt}{}{-0.10}{+0.10}}$ \\
GDS-7807 & 53.10476 & -27.8323 & 5.3956 & Not broad & AGN & 35.25 $\pm$ 10.20 &  $-2.16^{\genfrac{}{}{0pt}{}{-0.20}{+0.20}}$ \\
GDS-13176 & 53.12176 & -27.7976 & 5.9407 & Not broad & -- & 15.09 $\pm$ 6.06 &  $-2.13^{\genfrac{}{}{0pt}{}{-0.06}{+0.06}}$ \\
GDS-13197 & 53.13492 & -27.7727 & 6.342 & Not broad & Potential AGN & 9.62 $\pm$ 2.98 & $-2.10^{\genfrac{}{}{0pt}{}{-0.06}{+0.05}}$ \\
GDS-13704 & 53.12654 & -27.8181 & 5.9296 & AGN & AGN & 14.70 $\pm$ 6.85 & $-2.68^{\genfrac{}{}{0pt}{}{-0.13}{+0.13}}$ \\
GDS-113719 & 53.12429 & -27.7974 & 5.445 & Not broad & Potential AGN & 2.45 $\pm$ 1.46 & $-2.38^{\genfrac{}{}{0pt}{}{-0.12}{+0.11}}$ \\
GDS-202208 & 53.16407 & -27.7997 & 5.4489 & Not broad & SFG & 10.08 $\pm$ 2.32 &  $-1.98^{\genfrac{}{}{0pt}{}{-0.03}{+0.04}}$ \\
GDS-210003 & 53.13184 & -27.7738 & 5.7836 & Not broad & Potential AGN & 16.32 $\pm$ 3.28 & $-2.41^{\genfrac{}{}{0pt}{}{-0.13}{+0.12}}$ \\
GDS-212506 & 53.15584 & -27.7667 & 5.3503 & Not broad & Potential AGN & 16.29 $\pm$ 8.27 &  $-1.64^{\genfrac{}{}{0pt}{}{-0.16}{+0.16}}$ \\
GDS-30080593 & 53.14885 & -27.8097 & 5.9602 & Not broad & -- & 50.21 $\pm$ 25.15 &  $-1.99^{\genfrac{}{}{0pt}{}{-0.55}{+0.52}}$ \\
JADES-9104 & 189.2453 & 62.25253 & 6.8199 & Not broad & -- & 26.69 $\pm$ 13.47 & $-3.12^{\genfrac{}{}{0pt}{}{-0.39}{+0.36}}$ \\
JADES-721 & 189.1153 & 62.2341 & 5.1799 & Not broad & -- & 21.16 $\pm$ 18.47 &  $-2.49^{\genfrac{}{}{0pt}{}{-0.26}{+0.25}}$ \\
JADES-38147 & 189.2707 & 62.14842 & 5.8778 & AGN & -- & 22.83 $\pm$ 10.98 & $-1.07^{\genfrac{}{}{0pt}{}{-0.33}{+0.32}}$ \\
JADES-167639 & 53.09188 & -27.8807 & 5.5553 & AGN & -- & 19.38 $\pm$ 11.16 &  $-2.14^{\genfrac{}{}{0pt}{}{-0.31}{+0.29}}$ \\
RUBIES-16915 & 215.0796 & 52.93826 & 5.06 & AGN & AGN & 13.23 $\pm$ 3.31 & $-1.78^{\genfrac{}{}{0pt}{}{-0.11}{+0.11}}$ \\
RUBIES-45472 & 214.791 & 52.79478 & 5.2854 & Not broad & -- & 27.98 $\pm$ 16.23 &  $-1.62^{\genfrac{}{}{0pt}{}{-0.30}{+0.30}}$ \\
RUBIES-64705 & 214.9143 & 52.92779 & 5.2959 & Not broad & -- & 140.58 $\pm$ 46.46 & $-0.89^{\genfrac{}{}{0pt}{}{-0.87}{+0.88}}$ \\
RUBIES-46803 & 214.868 & 52.85372 & 5.1865 & Not broad & Potential AGN & 17.20 $\pm$ 8.06 &  $-1.21^{\genfrac{}{}{0pt}{}{-0.29}{+0.29}}$ \\
RUBIES-49140 & 214.8922 & 52.87741 & 6.6832 & AGN & AGN & 26.01 $\pm$ 11.99 &  $-1.79^{\genfrac{}{}{0pt}{}{-0.25}{+0.26}}$ \\
RUBIES-970128 & 214.881 & 52.89121 & 6.4858 & Not broad & AGN & 26.10 $\pm$ 12.43 & $-2.30^{\genfrac{}{}{0pt}{}{-0.24}{+0.24}}$ \\
RUBIES-15072 & 34.26858 & -5.28956 & 5.1948 & Not broad & AGN & 32.44 $\pm$ 13.04 & $-2.35^{\genfrac{}{}{0pt}{}{-0.37}{+0.37}}$ \\
RUBIES-174752 & 34.20581 & -5.1005 & 6.0426 & AGN & AGN & 23.39 $\pm$ 10.50 & $-2.06^{\genfrac{}{}{0pt}{}{-0.13}{+0.12}}$ \\
RUBIES-127820 & 34.30157 & -5.16619 & 6.9915 & No Halpha & -- & 121.34 $\pm$ 113.85 &  $-3.47^{\genfrac{}{}{0pt}{}{-0.62}{+0.76}}$ \\
RUBIES-125637 & 34.27878 & -5.16912 & 6.4192 & Not broad & -- & 45.13 $\pm$ 27.71 & $-1.49^{\genfrac{}{}{0pt}{}{-0.31}{+0.34}}$ \\
RUBIES-146730 & 34.32501 & -5.14008 & 5.2244 & Not broad & Potential AGN & 12.07 $\pm$ 5.65 & $-2.23^{\genfrac{}{}{0pt}{}{-0.18}{+0.18}}$ \\
RUBIES-35376 & 34.26532 & -5.2541 & 5.3161 & Not broad & -- & 16.65 $\pm$ 8.00 & $-2.18^{\genfrac{}{}{0pt}{}{-0.26}{+0.26}}$ \\
RUBIES-174222 & 34.4277 & -5.10124 & 6.08 & Not broad & -- & 284.16 $\pm$ 223.65 & $-0.35^{\genfrac{}{}{0pt}{}{-2.43}{+2.72}}$ \\
RUBIES-37427 & 34.50494 & -5.25798 & 6.958 & AGN & -- & 40.83 $\pm$ 20.23 & $-1.01^{\genfrac{}{}{0pt}{}{-0.52}{+0.53}}$ \\
UNCOVER-11254 & 3.580446 & -30.405 & 6.8727 & Not broad & Potential AGN & 11.59 $\pm$ 5.36 &  $-2.62^{\genfrac{}{}{0pt}{}{-0.11}{+0.11}}$ \\
UNCOVER-33673 & 3.600114 & -30.3658 & 6.7734 & Not broad & AGN & 53.02 $\pm$ 31.76 &  $-2.05^{\genfrac{}{}{0pt}{}{-0.93}{+0.95}}$ \\
UNCOVER-41225 & 3.533996 & -30.3533 & 6.7695 & AGN & AGN & 21.44 $\pm$ 9.50 & $-2.61^{\genfrac{}{}{0pt}{}{-0.51}{+0.51}}$ \\
UNCOVER-51076 & 3.553695 & -30.3301 & 5.9289 & Not broad & Potential AGN & 13.03 $\pm$ 2.06 &  $-1.13^{\genfrac{}{}{0pt}{}{-0.38}{+0.38}}$ \\
CAPERS-35805 & 150.0554 & 2.291588 & 6.532 & AGN & Potential AGN & 18.00 $\pm$ 8.28 & $-2.16^{\genfrac{}{}{0pt}{}{-0.10}{+0.10}}$ \\
CAPERS-52661 & 150.1418 & 2.193836 & 6.764 & AGN & AGN & 31.09 $\pm$ 12.17 &  $-1.623^{\genfrac{}{}{0pt}{}{-0.16}{+0.15}}$ \\
CAPERS-17158 & 150.1453 & 2.384542 & 5.355 & Not broad & Potential AGN & 9.83 $\pm$ 4.70 &  $-2.21^{\genfrac{}{}{0pt}{}{-0.15}{+0.15}}$ \\
CAPERS-39643 & 150.1077 & 2.271908 & 6.047 & Not broad & -- & 23.75 $\pm$ 11.54 & $-2.38^{\genfrac{}{}{0pt}{}{-0.27}{+0.26}}$ \\
CAPERS-49388 & 150.1393 & 2.214673 & 5.507 & Not broad & AGN & 7.24 $\pm$ 2.13 & $-2.21^{\genfrac{}{}{0pt}{}{-0.05}{+0.05}}$ \\
CAPERS-39948 & 150.1043 & 2.270252 & 5.298 & Not broad & AGN & 34.88 $\pm$ 16.28 &  $-1.79^{\genfrac{}{}{0pt}{}{-0.42}{+0.42}}$ \\
CAPERS-49845 & 150.1399 & 2.212099 & 5.126 & Not broad & AGN & 20.41 $\pm$ 6.96 & $-2.17^{\genfrac{}{}{0pt}{}{-0.15}{+0.15}}$ \\
CAPERS-11227 & 150.1195 & 2.418575 & 5.24 & Not broad & -- & 48.03 $\pm$ 16.06 &  $-2.96^{\genfrac{}{}{0pt}{}{-0.53}{+0.52}}$ \\
CAPERS-101195 & 150.0616 & 2.269404 & 5.989 & Not broad & AGN & 51.59 $\pm$ 22.30 &  $-3.02^{\genfrac{}{}{0pt}{}{-0.63}{+0.65}}$ \\
CAPERS-9244 & 214.9857 & 52.95623 & 5.203 & AGN & AGN & 24.24 $\pm$ 9.28 &  $-2.17^{\genfrac{}{}{0pt}{}{-0.10}{+0.10}}$ \\
CAPERS-19346 & 214.8868 & 52.82463 & 6.193 & Not broad & AGN & 22.23 $\pm$ 7.90 & $-2.22^{\genfrac{}{}{0pt}{}{-0.16}{+0.16}}$ \\
CAPERS-21137 & 214.8773 & 52.80983 & 5.679 & Not broad & -- & 22.84 $\pm$ 10.22 &  $-2.08^{\genfrac{}{}{0pt}{}{-0.25}{+0.23}}$ \\
CAPERS-80959 & 214.935 & 52.9326 & 6.517 & Not broad & AGN & 162.26 $\pm$ 106.70 &  $-3.33^{\genfrac{}{}{0pt}{}{-0.96}{+1.43}}$ \\
CAPERS-22908 & 214.8713 & 52.79732 & 5.687 & Not broad & -- & 18.053 $\pm$ 6.84 & $-2.81^{\genfrac{}{}{0pt}{}{-0.49}{+0.47}}$ \\
CAPERS-206147 & 214.8607 & 52.78599 & 5.68 & Not broad & Potential AGN & 13.71 $\pm$ 4.83 & $-2.03^{\genfrac{}{}{0pt}{}{-0.15}{+0.15}}$ \\
CAPERS-12328 & 214.813 & 52.81698 & 5.276 & Not broad & -- & 23.67 $\pm$ 11.95 &  $-1.96^{\genfrac{}{}{0pt}{}{-0.21}{+0.22}}$ \\
CAPERS-27271 & 214.8782 & 52.78325 & 5.283 & Not broad & Potential AGN & 16.03 $\pm$ 3.86 &  $-2.41^{\genfrac{}{}{0pt}{}{-0.14}{+0.14}}$ \\
CAPERS-62484 & 214.8562 & 52.81844 & 6.281 & Not broad & -- & 82.38 $\pm$ 54.93 &  $-1.08^{\genfrac{}{}{0pt}{}{-1.45}{+1.50}}$ \\
CAPERS-22230 & 214.99 & 52.88462 & 5.0878 & Not broad & AGN & 19.68 $\pm$ 9.07 & $-2.35^{\genfrac{}{}{0pt}{}{-0.19}{+0.20}}$ \\
CAPERS-41455 & 215.0059 & 52.8749 & 6.913 & Not broad & Potential AGN & 51.29 $\pm$ 25.56 & $-2.42^{\genfrac{}{}{0pt}{}{-0.33}{+0.32}}$ \\
\hdashline
CEERS-1334 & 214.7684 & 52.71764 & 5.4981 & AGN & Potential AGN & 17.19 $\pm$ 4.69 &  $-1.67^{\genfrac{}{}{0pt}{}{-0.14}{+0.14}}$ \\
GDS-5093 & 53.12417 & -27.8745 & 5.3981 & Not broad & Potential AGN & 13.47 $\pm$ 3.07 & $-2.10^{\genfrac{}{}{0pt}{}{-0.10}{+0.10}}$ \\
j0252m0503-376 & 43.06545 & -5.01028 & 6.7112 & Not broad & AGN & 33.40 $\pm$ 21.40 &  $-2.62^{\genfrac{}{}{0pt}{}{-0.49}{+0.49}}$ \\
macsj0647-1120 & 101.9015 & 70.19888 & 5.2358 & Not broad & -- & 24.13 $\pm$ 12.66 &  $-2.26^{\genfrac{}{}{0pt}{}{-0.31}{+0.30}}$ \\
PEARLS-184 & 265.0099 & 69.01782 & 6.2769 & Not broad & -- &  4.08 $\pm$ 1.43 & $-2.12^{\genfrac{}{}{0pt}{}{-0.04}{+0.04}}$ \\
RUBIES-157664 & 34.20792 & -5.12481 & 5.3122 & Not broad & AGN & 282.41 $\pm$ 266.24 &  $-1.35^{\genfrac{}{}{0pt}{}{-0.82}{+0.90}}$
\end{longtable}
}

\begin{table*}[!ht]
\centering
\caption{UV and optical emission-line ratios and equivalent widths for stacked samples.}
\label{tab:stack_all_lines}
\begin{tabular}{l c c c}
\toprule
& \textbf{WEAK stack} & \textbf{STRONG stack} & \textbf{NO \ion{C}{III}] stack} \\
\midrule
\multicolumn{4}{l}{\textit{Flux ratios}} \\
\ion{C}{IV}/\ion{He}{II}      & $0.403 \pm 0.400$   & $2.6 \pm 0.8$  & $<1.14$ \\
\ion{C}{III}]/\ion{He}{II}              & $4.4 \pm 1.4$  & $5.6 \pm 1.6$  & $4.6 \pm 1.6$ \\
{[\ion{Ne}{III}]/[\ion{O}{II}]}  & $1.3 \pm 0.1$  & $1.0 \pm 0.1$  & $0.90 \pm 0.07$ \\
{[\ion{O}{III}]/H$\beta$}            & $6.6 \pm 0.4$ & $5.9 \pm 0.3$ & $5.3 \pm 0.1$ \\
\midrule
\multicolumn{4}{l}{\textit{Equivalent Width}} \\
\ion{C}{IV} $\lambda\lambda1548,1551$      & $0.81 \pm 0.77$   & $12.6 \pm 1.7$  & $<0.92$ \\
\ion{C}{III}] $\lambda1908$                & $13.5 \pm 0.6$  & $43.5 \pm 3.6$  & $5.7 \pm 0.3$ \\
\bottomrule
\end{tabular}
\end{table*}

\end{appendix}

\end{document}